\documentclass{article}
\usepackage[paper=a4paper]{geometry}
\usepackage{graphicx} 
\usepackage{amsmath}
\usepackage{amssymb}
\usepackage{booktabs}
\usepackage{multirow}
\usepackage{natbib}
\usepackage{comment}
\usepackage{xcolor}
\usepackage{colortbl}
\usepackage{caption}
\usepackage{subcaption}
\usepackage{hyperref}
\usepackage{authblk}
\usepackage{algorithm}%
\usepackage{algorithmicx}%
\usepackage{algpseudocode}%
\usepackage{setspace}
\usepackage{enumitem}
\usepackage{adjustbox}
\usepackage[english]{babel}

\def \xvec {\text{\boldmath$x$}}    
\def \yvec {\text{\boldmath$y$}}    \def \mY {\text{\boldmath$Y$}}

\def \betavec         {\text{\boldmath$\beta$}}

\def \thetavec        {\text{\boldmath$\theta$}}

\title{Enhanced variable selection for boosting sparser and less complex models in distributional copula regression}

\author[1]{Annika Strömer}
\author[2]{Nadja Klein}
\author[1]{Christian Staerk}
\author[3]{Florian Faschingbauer}
\author[1,4]{Hannah Klinkhammer}
\author[1]{Andreas Mayr}

\small{
\affil[1]{Department of Medical Biometrics, Informatics, and Epidemiology, University Hospital Bonn, Germany}
\affil[2]{Department of Statistics, Technische Universität Dortmund, Dortmund, Germany}
\affil[3]{Department of Obstetrics and Gynecology, University Hospital of Erlangen, Erlangen, Germany}
\affil[4]{Institute for Genomic Statistics and Bioinformatics,
University Hospital Bonn, Bonn, Germany}
}

\date{}
\begin{document}
\maketitle

\begin{abstract}
Structured additive distributional copula regression allows to model the joint distribution of multivariate outcomes by relating all distribution parameters to covariates. Estimation via statistical boosting enables accounting for high-dimensional data and incorporating data-driven variable selection, both of which are useful given the complexity of the model class. However, as known from univariate (distributional) regression, the standard boosting algorithm tends to select too many variables with minor importance, particularly in settings with large sample sizes, leading to complex models with difficult interpretation. 
To counteract this behavior and to avoid selecting base-learners with only a negligible impact, we combined the ideas of probing, stability selection and a new deselection approach with statistical boosting for distributional copula regression. In a simulation study and an application to the joint modelling of weight and length of newborns, we found that all proposed methods enhance variable selection by reducing the number of false positives. However, only stability selection and the deselection approach yielded similar predictive performance to classical boosting. Finally, the deselection approach is better scalable to larger datasets and led to a competitive predictive performance, which we further illustrated in a genomic cohort study from the UK Biobank by modelling the joint genetic predisposition for two phenotypes.
\end{abstract}

\section{Introduction}\label{sec1}
Statistical boosting, an iterative, sequential fitting algorithm for statistical models originating from machine learning~\citep{Freund1990}, has gained increasing interest as an alternative to classical (penalized) maximum likelihood estimation (PLME) or Bayesian inference. Since boosting is well-suited for high-dimensional and complex data problems, it is also a useful tool for distributional regression, where the number of candidate models is typically large. Distributional regression generally has the aim of estimating complete conditional distributions of a quantity of interest as a function of covariates~\citep[see e.g.,][for a recent review]{Kle2024}. A convenient framework for univariate distributional regression is the class of generalized additive models for location, scale and shape \citep[GAMLSS;][]{gamlss}, which allow relating each distribution parameter of a parametric response distribution to covariates. However, in the classical PLME-based implementation, the response is restricted to be univariate and the complexity of the predictors is limited due to numerical instabilities when it comes to selecting smoothing parameters for regularization. While Bayesian estimation of GAMLSS based on Markov chain Monte Carlo simulations \citep{KleKneKlaLan2015, bamlss} allows to overcome the latter issues, it is notoriously slow when the number of observations~$n$ and/or the number of covariates~$p$ is large. In such scenarios statistical boosting is particularly beneficial as also  demonstrated in various applications and extensions of the original boosting algorithm~\citep[see e.g.,][]{staerk2021randomized, klink2023PRS}. 

In this paper, we are not only concerned with situations, where either $n$ or $p$ is large or where even $p\gg n$, but particularly when the outcome~$Y$ is multivariate. By modelling interrelated outcomes together, we can gain a better understanding of the relationship and identify relevant factors affecting their association. An example is the consideration of multiple phenotypes from deeply phenotyped cohort studies in genetic epidemiology. 
For distributional regression modelling of multiple outcomes, one approach is to consider the joint parametric distribution. A popular alternative in this context are copulas, that offer increased flexibility by allowing the use of different marginals and different dependence structures through the copula function. This approach has been the subject of ongoing research and there is rich literature on copula modelling with regression data \citep[see e.g.,][for recent examples]{YanCza2022,nguyen2023power,VerFloMolKei2024}.

Statistical boosting was extended to multivariate distributional regression towards parametric distributions \citep{Stroemer2023} and also using copula \citep{Hans2022}. However, while this is done conceptually in these works, some practical aspects are still challenging. One challenge is that while boosting has an implicit variable selection mechanism, it often leads to relatively \emph{large} models, that is, models with many included covariates despite having small to negligible effects. This happens because the algorithm typically optimizes prediction accuracy without explicitly considering sparsity. This was also observed for boosting copula regression, where especially the sub-models for the location parameter did result in rather large numbers of selected covariates~\citep{Hans2022}. This behaviour is particularly undesirable in situations with a large number of candidate variables~$p$, where sparse and interpretable models are practically relevant. Therefore, variable selection is of great importance, not only in the context of boosting~\citep[see e.g.,][]{tian2024variable,keil2023considerations}. In distributional copula regression, a further interesting question that arises is how to decide in a data-driven manner if the overall complexity of the model could be reduced. In this context, the aim of statistical modelling is to select a model that is as complex as needed but also as simple as possible to facilitate efficient estimation and interpretability. For example, if certain distribution parameters like the association parameters do not depend on covariates, then simpler univariate models might also be suitable. Statistical boosting can help to answer this question.

To tackle these yet unaddressed practical challenges in boosting distributional copula regression, we incorporate three existing approaches for refining variable selection within this framework. All three approaches have been already proposed or extended to boosting, but have never been integrated into boosting multivariate distributional regression via copulas. This new combination aims to reduce the complexity of the model, particularly when dealing with high-dimensional data. The three considered existing approaches for enhanced variable selection are the following. (i) Stability selection~\citep{Meinshausen2010}, which has been extended to boosting univariate distributional regression~\citep{Thomas2018}; (ii) probing~\citep{Thomas2017}, which was proposed for boosting simple mean regression models, shifts the focus of early stopping from prediction accuracy directly to variable selection; and (iii) deselection~\citep{Deselection}, the newest approach, which pragmatically deselects base-learners that do not contribute enough to the overall model performance.

We initially investigate the performance of these three methods on simulated data, where the true data-generating process is known. Afterwards, we consider two real data applications: first, the joint modelling of the weight and length of newborns \citep[as in][]{Hans2022}, and second, the modelling of the joint genetic disposition towards continuous phenotypes based on a large cohort study (UK Biobank).

\section{Methods}\label{sec2}

\subsection{Boosting distributional copula regression}
In this section, we briefly review distributional copula regression models focusing on the bivariate case of two continuous outcomes and how statistical boosting algorithms can be applied for model estimation.  

\subsubsection{Distributional copula regression models}
A flexible modelling approach for the joint analysis of two continuous response variables $\mY=(Y_1, Y_2)^\top$ in terms of covariates are bivariate copula regression models, which describe the dependence structure through a copula \citep{Nelsen2006}. 
According to Sklar's theorem, the joint conditional cumulative distribution function (CDF) of two responses given covariate information $\xvec$ can be written as
\begin{equation*}
    F(y_1, y_2 \mid\thetavec) = C\left[F_1(y_1 \mid\thetavec^{(1)}), F_2(y_2 \mid \thetavec^{(2)})\mid\thetavec^{(c)}\right],
\end{equation*}
where $F_1(\cdot|\,\thetavec^{(1)})$ and $F_2(\cdot|\,\thetavec^{(2)})$ are the marginal conditional CDFs of the two responses $Y_1=y_1$ and $Y_2=y_2$ which are uniformly distributed on $[0,1]$. The copula function  $C(\cdot,\cdot\,|\,\thetavec^{(c)})$ contains the information about the dependence structure between the two outcomes and is unique when the responses are continuous. The vector $\thetavec = \lbrace (\thetavec^{(1)})^T,(\thetavec^{(2)})^T,(\thetavec^{(c)})^T\rbrace^\top$ contains the model parameters $k = 1,\dots, K$ of the marginal distributions and the copula, whereby all components of $\thetavec\equiv\thetavec(\xvec)$ can be linked to a covariate vector via additive predictors and appropriate link functions. The representation of the joint conditional CDF via a copula allows the separation of the marginal distributions and the dependence structure; different copula functions allow different structures to be modeled. The Clayton copula, for example, can capture asymmetric dependence (so-called lower tail dependence) where the two responses show a stronger positive association for smaller values than for larger values. In our work we will focus on Gaussian, Clayton and Gumbel copulas~\citep[cf.][]{Nelsen2006} to represent no, lower and upper tail dependencies. 

The joint density $f(y_1,y_2\,|\,\thetavec)$ of a distributional copula regression model can be expressed via
\begin{align*}
    f(y_1,y_2\,|\,\thetavec)&= \frac{\partial^2 }{\partial F_1 \partial F_2} F(y_1,y_2\,|\,\thetavec)\\ &=c\left[F_1(y_1 \,|\,\thetavec^{(1)}), F_2(y_2 |\thetavec^{(2)})\,|\,\thetavec^{(c)}\right]\, 
    f_1(y_1\mid\thetavec^{(1)})\,f_2(y_2\mid\thetavec^{(2)}) ,
\end{align*}
where $f_1(\cdot\,|\,\thetavec^{(1)})$ and $f_2(\cdot\,|\,\thetavec^{(2)})$ are the marginal probability density functions and $c(\cdot,\cdot\mid\thetavec^{(c)})$ is the copula density of $C$. Based on our applications, the most relevant marginal distributions in this work are the log-logistic and the log-normal distributions. 

Finally, for a dataset of $n$ independent pairs $\lbrace(\yvec_i,\xvec_i)\rbrace_{i=1}^n$ of bivariate responses $\yvec_i=(y_{i1},y_{i2})^\top$ with covariate information $\xvec_i$, the joint log-likelihood function is given by
\begin{equation*}
    l(\thetavec) = \sum_{i=1}^n \log\left\lbrace c\left[F_1(y_{i1}\,|\,\thetavec^{(1)}),F_2(y_{i2}\,|\,\thetavec^{(2)})\,|\,\thetavec^{(c)}\right]\right\rbrace + \sum_{i =1}^n \sum_{d\in \{1,2\} } \log\left\lbrace f_d(y_{id}\,|\,\thetavec^{(d)})\right\rbrace.
\end{equation*}

\subsubsection{Structured additive predictors}
In distributional copula regression, each distribution parameter $\theta_k$, $k=1,\ldots,K$ is modelled via a structured additive predictor $\eta_k$~\citep{FahKneLan2004} with parameter-specific monotonic link functions $g_k$, such that $g_k(\theta_k)=\eta_k$ and $g_k^{-1}(\eta_k)=\theta_k$, where $g_k^{-1}$ is the inverse of $g_k$. The additive predictors $\eta_k$ depend on (possibly different) subsets of $\xvec$, 

  $$g_k (\theta_k) = \eta_k =  \beta_{0k} + \sum\limits_{j = 1}^{p_k} f_{jk}(\xvec),     \mbox{ for } k = 1,\dots, K,$$

where $\beta_{0k}$ are the intercepts and each $f_{jk}$, $j=1,\ldots,p_k$, represents functional effects of covariates $\xvec$.
The effects can be chosen in a flexible manner~\citep{FahKneLanMar2013}, for instance we incorporate linear and non-linear effects in Sections~\ref{Simulation} and~\ref{Application}. Linear effects can be represented by $f_{jk}(\xvec) = \xvec_{jk}^T \betavec_{jk}$ whereby $\xvec_{jk}$ is a covariate subset of $\xvec$ for parameter~$\theta_k$ and $\betavec_{jk}$ are the regression coefficients. Non-linear effects can be included using appropriate basis functions, such as B-splines.

\subsubsection{Estimation via model-based boosting}
Component-wise gradient boosting with regression-type base-learners, also referred to as \emph{statistical boosting}~\citep{mayr2014evolution}, originates from the gradient boosting approach of~\cite{Freund1990}, who translated the original concept from the machine learning literature to statistical modelling. Its basic idea is to iteratively minimize a pre-specified loss function~$\mathcal{L}$ by fitting so-called base-learners separately to the negative gradient~$\mathcal{L}$ and by then adding only a small amount of the ``best-fitting'' base-learner---that is, the base-learner that yields the steepest descent in the direction of the current gradient---to the overall regression predictor in each step of the boosting algorithm. Each base-learner usually represents one effect in the additive regression predictor (see~\cite{Hofner2014} for a detailed overview). In this way, the overall predictor is built sequentially, where more and more variables are selected the longer the algorithm runs, such that \emph{early stopping} yields implicit variable selection. In likelihood-based statistical boosting, the loss~$\mathcal{L}$ is the negative log-likelihood $l\equiv l(\thetavec)$, but more general functionals such as proper scoring rules are possible.

The boosting algorithm is a flexible alternative to classical estimation approaches. It has several practical advantages, such as dealing with high-dimensional data in which classical inferential methods are no longer applicable. As mentioned above, the algorithm performs data-driven variable selection, which is controlled by the number of boosting iterations $m_{\text{stop}}$~\citep{MayrHofnerSchmid2012}: Variables whose corresponding base-learner has never been selected until $m_{\text{stop}}$ is reached, are excluded from the final model. Therefore, the number of boosting iterations is the main tuning parameter and is typically optimized by cross-validation or resampling techniques. 
Another parameter of the algorithm is the fixed step length $\nu$, with which the best-fitting base-learner is multiplied before being included into the predictor. It is commonly fixed to $\nu = 0.1$~\citep{SCHMID2008298}. 

In the boosting approach for distributional copula regression~\citep{Hans2022}, all distribution parameters are modeled simultaneously by combining the properties of GAMLSS and the main features of statistical boosting. In every iteration, the partial derivatives $u_k = \partial l/ \partial\theta_k$ of the negative log-likelihood $l$ with respect to the different distribution parameters $\theta_k$ are calculated and each base-learner $h_{jk}\equiv f_{jk}(\xvec_{jk})$ is separately fitted to the gradient. Then, the best-fitting base-learner (and the corresponding update) for each distribution parameter is determined and compared across the different dimensions. Only the overall best-performing update is finally performed using a non-cyclic version of the algorithm~\citep{Thomas2018}. For more details on fitting distributional copula regression via boosting, we refer to \cite{Hans2022}.

\subsection{\mbox{Complexity reduction \& enhanced variable selection}}\label{Sec:VarSel}

In the following, we present different techniques for enhanced variable selection that we will integrate in our boosting distributional copula framework. Probing (Section~\ref{prob}) had been introduced to statistical boosting by~\cite{Thomas2017} and since then been applied or used as a benchmark approach for mean regression models with only one dimension \citep{staerk2021randomized, dikheel2022using} or joint models of time-to-event and longitudinal data \citep{griesbach2023variable}. Stability selection (Section~\ref{stabsel}) is a more general approach~\citep{Meinshausen2010} and has been introduced to boosting mean-regression models by~\cite{HofnerStablsel}, before the approach was extended to the context of univariate distributional regression~\citep{Thomas2018}. Deselection (Section~\ref{Desel}) is the most recent enhancement and was directly introduced for boosting mean regression as well as distributional regression~\citep{Deselection}. None of the three approaches have ever been extended towards boosting multivariate distributional regression or even to copula regression. In the process of integrating these enhanced variable selection approaches in our framework, we also allow for constant distribution parameters that do not depend on covariates. This is particularly attractive in copula regression, where for example copula parameters not depending on covariates reflect situations in which the dependence structure between the outcomes does not vary across observations with distinct feature values. This can lead to a substantial reduction in the complexity of the final model.

\subsubsection{Probing}\label{prob}
\textit{Probing} is based on the inclusion of random noise variables, so-called probes, to determine the stopping iteration by stopping when the algorithm starts selecting those (Algorithm~\ref{Alg:Probing}): First, randomly generated shuffled versions (probes) of the covariates are added to the original dataset. Second, a boosting model is fitted on the expanded dataset and the algorithm stops when the first probe is selected. The idea is that, in each iteration, the base-learner with the highest loss reduction is updated and the selection of a probe means that the best possible improvement is based on information known to be unrelated to the outcome. Because each parameter may depend on a potentially different set of variables, the randomly shuffled probes are simply added for each of the distribution parameters. In our model class, the distribution parameters may represent parameters of the marginal distributions or the copula. In each boosting iteration, a single base-learner is updated, i.e., the algorithm stops when the first probe is selected for any of the distribution parameters. While probing does not require optimizing the stopping criterion via computationally expensive cross-validation or resampling, it optimizes towards sparse models and does not maximize prediction performance. As a consequence, probing typically yields sparse models with strongly regularized predictor effects~\citep{Thomas2017}. 

\begin{algorithm}\setstretch{1.2}
\caption{Probing for boosting distributional copula regression.}
\label{Alg:Probing} 
\begin{algorithmic}[1]
    \State Shuffle probes $\Tilde{\xvec}_{jk}$ for each of the covariates  $\xvec_{jk}$ with $j = 1,\dots,p_k$ and $k=1,\dots,K$. 
   \State Perform boosting on the expanded set of variables $x_1,\dots, x_{p_k}, \Tilde{x}_1, \dots, \Tilde{x}_{p_k}$ for each distribution parameter $\theta_k, k=1,\dots,K$.
  \State Stop when the first probe $\Tilde{x}_{jk}$ of any distribution parameter is selected. 
  \State  Use final model from the previous iteration (containing only original variables). 
\end{algorithmic}
\end{algorithm}

\subsubsection{Stability selection}\label{stabsel}
A popular enhanced variable selection technique is \textit{stability selection}, which yields a stable set of covariates by repeated model fitting using subsamples of the original dataset~\citep{Meinshausen2010}. In the context of boosting, \cite{Thomas2018} introduced stability selection for boosted GAMLSS. As outlined in Algorithm~\ref{Alg:Stabsel}, the general idea is to draw $B$ random subsets of the data with size $\lfloor n/2 \rfloor$ of the original dataset and to fit separate boosting models for each subset. The boosting algorithm runs on each subset until a pre-specified number of covariates~$q$ have been selected. Every variable has a selection frequency defined by the fraction of subsets in which the variable $j$ was selected. If the selection frequency exceeds the threshold~$\pi_{\mathrm{thr}}$, the variable is considered stable and is included in the final model fit~\citep{HofnerStablsel}. 
Stability selection provides a sparse solution, controlling the number of false discoveries by defining an upper bound for the per-family error rate (PFER), i.e., the expected number~$\mathbb{E}(V)$ of noninformative variables included in the final model. The upper bound is given by $\mathbb{E}(V) \leq q^2 / ((2\pi_{\mathrm{thr}} - 1)p)$, where $p = \sum_{k=1}^K p_k$ is the total number of predictor variables and $q$ the number of selected variables. 

For practical use, the most important aspect is the choice of the parameters $q, \pi_{\mathrm{thr}}$ and PFER, whereby the PFER can be derived from the upper bound and visa versa. \cite{Meinshausen2010} state that the number of selected base-learners $q$ should be chosen sufficiently large concerning the informative variables, or at least as high as the number of informative variables, which, however, are usually unknown. The threshold~$\pi_{\mathrm{thr}}$ typically has only minor importance and should be in the range of $\pi_{\mathrm{thr}} \in (0.6,0.9)$ (a variable should be selected in more than half of the fitted models in order to be considered stable). 
\begin{algorithm}\setstretch{1.2}
\caption{Stability selection for boosting distributional copula regression.}
\label{Alg:Stabsel}
\begin{algorithmic}[1]
    \For{$b = 1,\dots,B$}
            \State Select a random subset from the data of size $\lfloor n/2  \rfloor$.
            \State Fit a boosting model until $q$ base-learner are selected. 
        \EndFor
    \State Compute the relative selection frequencies per base-learner $\hat{\pi}_j = \frac{1}{B} \sum I_{j\in \hat{S}_b}$, where $\hat{S}_b$ denotes the set of selected base-learner. 
    \State Select the stable set of base-learner $\hat{S}_{\rm{stable}}:= \{j: \hat{\pi}_j \geq \pi_{\rm{thr}}\}$.
    \State Fit a boosting model with the stable set of base-learners.
\end{algorithmic}
\end{algorithm}

\subsubsection{Deselection of base-learners}\label{Desel}
Another approach to encourage variable selection and sparsity is to deselect and remove base-learners with a negligible impact on the model's predictive performance. The general idea is to start with a classical boosted model tuned by cross-validation or resampling techniques. Then, the base-learners that were selected but only have a minor impact on the model are identified and deselected.
Afterwards, the model is boosted again with the remaining variables. This idea was introduced by~\cite{Deselection} for univariate GAMLSS and is now extended to distributional copula regression. The importance of a base-learner is based here on the risk reduction and can be defined for base-learner $j$ after $m_{\rm{stop}}$ boosting iterations with 
\begin{equation*}\label{eq:RiskRed}
R_j = \sum_{m=1}^{m_{\rm{stop}}}I(j = j^{*[m]}) (r^{[m-1]} - r^{[m]}), \quad j = 1,\dots,  p,
\end{equation*}
where $I$ denotes the indicator function and $j^{*[m]}$ is the selected base-learner in iteration $m$. Furthermore,  $r^{[m-1]} - r^{[m]}$ represents the risk reduction in iteration $m$, for risks $r^{[m]}$ and $r^{[m-1]}$ at iterations $m$ and $m-1$, respectively. Note that in the case of distributional copula regression, all distribution parameters are considered together and each parameter $\theta_k, k=1,\dots, K$ can depend on a different number of variables~$p_k$. Here, we do not distinguish between the different parameters, such that $p = \sum p_k$.

For a given threshold $\tau \in (0,1)$, we deselect base-learner $j$ if
	$$R_{j} <  \tau \cdot (r^{[0]} - r^{[m_{\rm{stop}}]}),$$
where $r^{[0]} - r^{[m_{\rm{stop}}]}$ represents the total risk reduction and $R_{j}$ denotes risk reduction attributable to  base-learner $j$. In other words, only base-learners whose contribution $R_j$ to the total risk reduction is larger than
the relative $\tau$ threshold~\citep[e.g., 1\%][]{Deselection} will remain in the model after the deselection step.

\begin{algorithm}\setstretch{1.2}
\caption{Deselection for boosting distributional copula regression.}
\label{Alg:Desel} 
\begin{algorithmic}[1]
\State Initial boosting: \\
              [] Tune $m_{\rm{stop}}$ based on cross-validation or resampling techniques (early stopping).
\State Deselection: \\
              [] Identify the base-learners with minor impact on the risk reduction according to $R_j < \tau \cdot (r^{[0]} - r^{[m_{\rm{stop}}]})$ and remove them from the model.
\State Final boosting: \\
          [] Boost again with the remaining variables and the $m_{\rm{stop}}$ of Step 1.
\end{algorithmic}
\end{algorithm}

\subsubsection{Illustration of the different approaches}

\begin{figure}[t!]\centering
    \includegraphics[width=\textwidth]{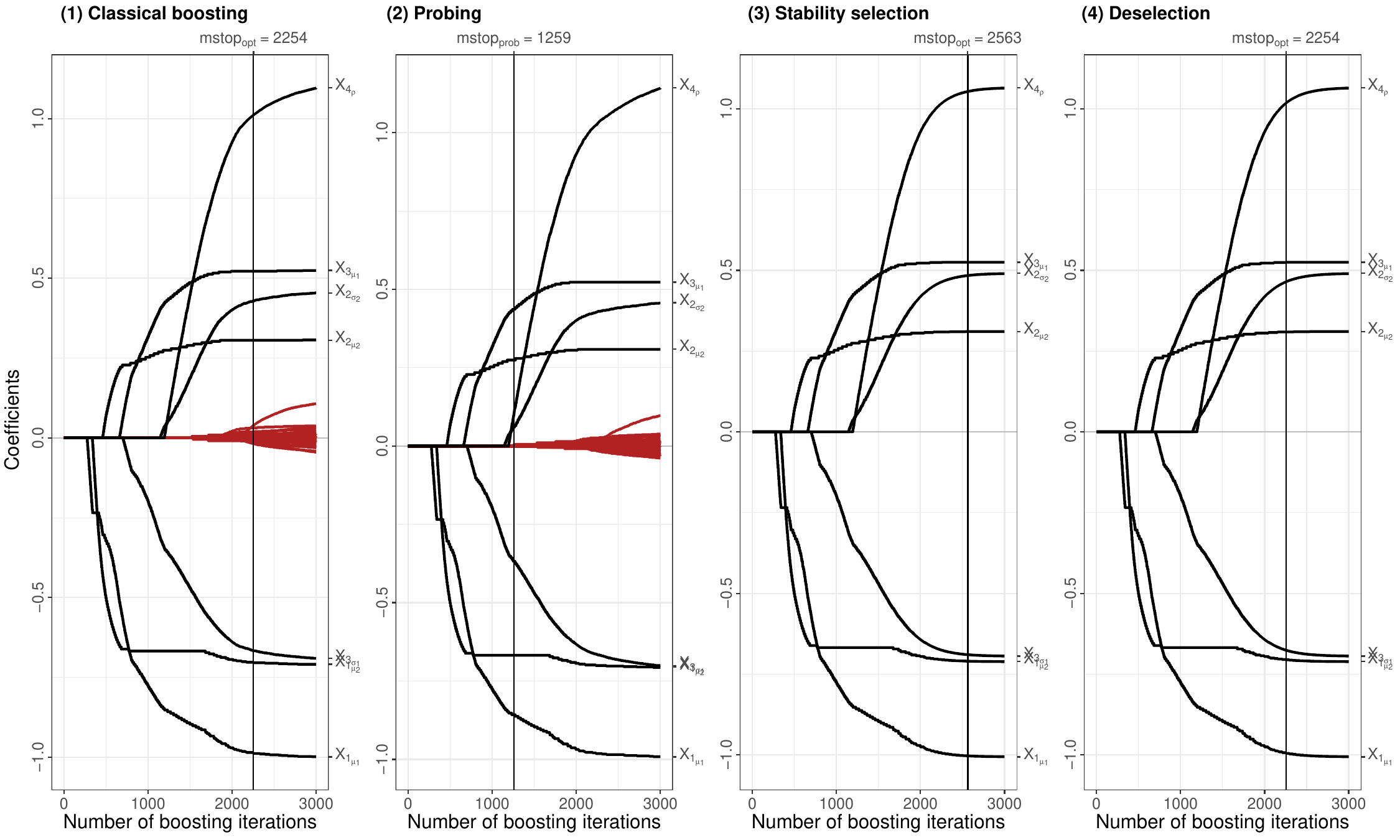}
    \caption{The resulting coefficient paths along the number of boosting iterations for a simulated example (for more details see the supplement: Section A) for the classical boosting, probing, stability selection and the deselection approach (1\%). The coefficient paths of the informative variables are colored in black, the noninformative in red. The intercept was removed for clarity. For stability selection and deselection, only the final model is plotted. 
    }\label{Fig:Method}
\end{figure}

Figure~\ref{Fig:Method} displays the coefficient paths resulting from the classical boosted copula regression and the final models after applying the different approaches for reducing the model complexity on simulated data (more details on this example can be found in the supplement: Section A). Overall the coefficient paths of the different approaches yield similar final models. Applying probing leads to earlier stopping than the classical model with a stopping iteration at 1259 iterations. Therefore, the effect estimates are shrunken and fewer variables are included in the model (all informative but also one noninformative variable). As described in Section~\ref{prob}, the shrinkage of the effect estimates might not be optimal for predictive performance. The resulting model for stability selection is shown in the third plot, with selection frequencies across the $B$ subsets for the different base-learners provided in the supplement (Figure~A1). The performance of stability selection depends strongly on the choice of the parameters, here we choose $q =20$ and $\text{PFER}=5$, but for example smaller $q$ and PFER would lead to worse results as most informative variables would not be included, leading to poorer predictive performance.  

The deselection approach with a threshold value of 1\% is similar to stability selection. Stable covariates are the ones with the highest risk reduction in the deselection approach. The final deselection model contains also here only the informative variables with the same number of boosting iterations as the classical model and similar coefficient estimates. The corresponding risk reduction for the different variables can be found in Figure~A2 in the supplement with different threshold values (0.1\% and 1\%). 
Higher threshold values would lead to the elimination of informative variables, whereby for smaller values such as 0.1\% (dotted line in the risk reduction plot Figure~A2 in the supplement) noninformative variables would remain in the model. 


\subsection{Computational details and implementation}
Boosting for distributional copula regression is implemented via the \textsf{R} package \textbf{gamboostLSS}. For tuning of $m_{\text{stop}}$, cross-validation, resampling techniques or evaluation on a single test dataset can be used. The process is facilitated using a provided function that directly works on the model object. 

From a computational and implementation perspective, probing can be very easily utilized, because no computationally intensive techniques for optimizing the stopping iteration and no additional tuning parameters are required. Stability selection for copula regression can be realized using the fitted boosting model and the \texttt{stabs()} function in the package~\texttt{gamboostLSS}. One needs to specify two of the parameters beforehand, the number of base-learners~$q$, the per-family error-rate and the threshold~$\pi_{\rm{thr}}$. The function returns the stable set of base-learner for each distribution parameter. To obtain the final model, one can again run a boosting model with only these stable base-learners. 
Moreover, the function encompasses various options for assumptions. It is important to note that the described approach in Section~\ref{stabsel} does not involve any additional assumptions (\texttt{assumption = "none"}).

The implementation of the deselection approaches is available at GitHub \url{https://github.com/AnnikaStr/ComplRedBoostCop} and is accessed with the \texttt{DeselectBoost()} function, which requires a boosting model with early stopping and the specification of an appropriate threshold value (e.g., 1\%). The re-fitting of the model with the remaining base-learners to obtain the final model is already included in the function.

\section{Simulations}\label{Simulation}
To evaluate the performance of the different approaches for reducing the model complexity of boosted bivariate distributional copula regression models, we conducted a simulation study. We compared probing, stability selection and the deselection of base-learners with a focus on their variable selection properties, the prediction performance and runtime. Our specific objectives were to determine: (i) Can the variable selection approaches identify the truly informative variables while decreasing the number of false positives? (ii) How do the approaches perform in comparison to each other? (ii) Can the complexity of the model be reduced by simplifying complete additive predictors for distribution parameters to an intercept?

A detailed description of the simulation design of the following scenarios can be found in the supplement: Section B. Furthermore, here we provide a descriptive summary of the simulation results, while detailed numerical results can be also found in the supplement: Section B. All code to reproduce the results can be found on GitHub~\url{https://github.com/AnnikaStr/ComplRedBoostCop}.  The simulations were conducted in \texttt{R} using the add-on package \textbf{gamboostLSS} for estimating the copula regression models. The \textbf{copula} and \textbf{gamlss} packages were used for data generation.

\subsection{Simulation design}
To investigate these questions, we considered three different bivariate scenarios for continuous outcomes, each with five distribution parameters (marginal means $\mu_1$ and $\mu_2$, marginal variances $\sigma_1^2$ and $\sigma_2^2$, and association parameter $\rho$): 

\vspace{0.2cm}
\begin{enumerate}[label=\textbf{Scenario~\Alph*},ref={Scenario~\Alph*},wide = 0pt,itemsep=.2cm]
\item\label{ModelA}
Same simulation setup as in~\cite{Hans2022} with four informative variables $x_1, \dots, x_4$. Cubic P-splines with 20 equidistant knots and a second-order difference penalty were included as base-learners. The log-normal and log-logistic distribution were used as marginal distributions. 
\item\label{ModelB} 
Modification of Scenario A: $\sigma_1$ does not depend on explanatory variables. 
\item\label{ModelC} More informative variables: Ten informative variables for each distribution parameter $p_k = 10, k = 1,\dots,5$ with $x_1, \dots, x_{50}$ with Gaussian marginal distributions for both outcomes. The base-learners correspond to simple linear models.
\end{enumerate}

A total of 100 simulation runs were performed for each simulation setting. For each scenario, $n=1000$ observations were considered, where the covariates $x_1, \dots, x_p$ were independently drawn from a uniform distribution on $(-1,1)$. The simulations cover a low-dimensional case ($p<n$) with $p = 20$ variables for~\ref{ModelA} and~\ref*{ModelB} and $p = 200$ for~\ref{ModelC}. Furthermore, a high-dimensional case ($p>n$) with $p = 1000$ variables was investigated for each scenario. The Gaussian, the Clayton and the Gumbel copula were considered. For fitting the models, all covariates were considered for each distribution parameter simultaneously. The stopping iteration $m_{\rm{stop}}$ was optimized by minimizing the empirical risk on an additional validation dataset with 1500 observations. For all simulations, the step length of the boosting algorithm was set to a fixed value of $\nu = 0.01$. 
For the deselection approach, we specified the threshold parameter $\tau$ with 0.1\% and 1\%. For stability selection, the number of variables to be included in the model was set as $q =20$ and the per-family error-rate was chosen to be $\mathrm{PFER} = 5$. Note that due to the high computational cost, stability selection could not be applied for the high-dimensional settings.

To evaluate the prediction performance we used multivariate proper scoring rules, namely the negative log-likelihood and the energy score. The latter generalizes the continuous ranked probability score for multivariate quantities~\citep{Gneiting2008AssessingPF}. 

\subsection{Summary of simulation results}
In~Scenarios A and B, the informative variables were correctly selected for each distribution parameter by classical boosting, while noninformative variables were included mainly for the mean parameters (Supplement: Section B.1 and B.2). However, in~\ref{ModelC} not all informative variables were selected for the dependence parameter and many noninformative variables were included for the mean and scale parameters (see the supplement: Section B.3). 
In comparison, probing, stability selection and deselection led to much sparser models in both low- and high-dimensional cases.
Specifically, in Scenarios A and B, the final models contained all informative variables except for probing, which occasionally did not contain the informative variable for the dependence parameter~$\rho$. Stability selection also did not select the informative variable for $\rho$ from time to time in Scenarios A and B using a Clayton copula. With the deselection approach, all informative variables remained in the model. The fewest false positives were obtained when $\tau$ was set to 1\%, almost completely eliminating them. With a threshold value of 0.1\% also many noninformative variables were excluded but to a lower extent than with 1\% (see the supplement: Section B.1 and B.2). 

In~\ref{ModelC}, only for the Gaussian copula all informative variables were selected by classical boosting; for the other copulas, it was already difficult to select all true positives for the dependence parameter. Most false positives were included for the mean and scale parameters. 
The resulting models for probing and stability selection for~\ref{ModelC} had difficulties in selecting the informative variables. The average number of true positives is relatively low for both approaches. 
The deselection approach with a threshold value of 0.1\% only slightly influenced the average number of true positives. For all other parameters, the informative variables remained in the model in every simulation run. The number of noninformative variables were considerably reduced but there are still false positives left in the model. A higher threshold value would lead to a higher decrease in false positives but also to a reduction of correctly identified informative variables (see the supplement: Section B.3).

For the predictive performance on test data, evaluated with the negative log-likelihood and the energy score (smaller values are better), the deselection approach as well as stability selection had a comparable predictive performance and led to an improvement in the negative log-likelihood compared to the classical boosting for~\ref{ModelA} and \ref{ModelB}. For the energy score, the approaches resulted in similar values. Only probing showed a worse performance compared to the classical boosted model for the negative log-likelihood and the energy score (see the supplement: Section B.1 and B.2). 
For \ref{ModelC}, probing and stability selection led to a worse predictive performance due to the exclusion of informative variables. The deselection approach yielded an improvement in the negative log-likelihood for a threshold value of 0.1\% and provided comparable performance to the classical approach regarding the energy score (see the supplement: Section B.3). 

Overall probing yielded the smallest runtime, as there is no need for an additional optimization of the stopping iteration. Due to its second boosting step, the deselection approach took slightly longer than the classic approach ($\approx$ 1--2 minutes). Stability selection had the longest runtime because $B$ boosting models have to be fitted on the subsamples. 

\section{Real data illustrations}\label{Application}

\subsection{Analysis of fetal ultrasound data}
Motivated by the analysis of fetal ultrasound data using boosted copula regression of~\cite{Hans2022}, which resulted in rather large sub-models for the different distribution parameters, we examined and compared the variable selection and the predictive performance of this analysis with the models resulting from the enhanced variable selection techniques introduced in Section~\ref{Sec:VarSel}.
The considered dataset was collected from 2006 to 20016 at the Department of Obstetrics and Gynecology of the Erlangen University Hospital and contains $6103$ observations and $36$ variables, including sonographic variables, e.g., abdominal anteroposterior diameter, abdominal transverse diameter, the interaction between these sonographic variables, and clinical variables, e.g., weight, height and body-mass-index (BMI) of the mother. For more details on the data, we refer to~\citep{Faschingbauer2016}. 

The response variables of interest are the birth length and weight, which were modeled via copula regression with log-logistic marginal distributions and the Gaussian copula. We split the dataset into a training dataset with $n = 4103$ observations and a test dataset for evaluation with $2000$ observations. The step length was set to $\nu  = 0.01$ and the stopping iteration was optimized by 10-fold cross-validation. All variables were considered for each distribution parameter. For continuous variables, cubic P-splines with 20 equidistant knots, a second-order difference penalty and 4 degrees of freedom were used as base-learners. Sex of the fetus and gestational diabetes were included via linear base-learners. Furthermore, we applied a gradient stabilization to ensure comparable gradients for the distribution parameters~\citep{Hofner2014}.
For the deselection approach, threshold values of 0.1\% and 1\% were considered. The parameters for stability selection were specified as $q=20$ for the number of variables to be included in the model and $\text{PFER} = 5$ for the per-family error-rate.

\begin{table}[t!]
    \centering
    \begin{tabular}{l|ccccccc}
    \toprule
     Method & $\mu_1$ & $\sigma_1$ & $\mu_2$ &  $\sigma_2$ & $\rho$ & $-$Log-Lik & $m_{\rm{stop}}$\\ \midrule
     classic & 33 & 15 & 30  & 13 & 9 & 4156.58 & 5536\\
     deselection 0.1\% & 6 & 9 & 8 & 7 & 4 & 4184.07 & 5536 \\ 
     deselection 1\% & - & 5 & 2 & 4 & 1 & 4843.41 & 5536 \\
     probing & 9 & 9 & 8  & 9 & 5 & 4654.58 & 808 \\ 
     stability selection & 5 & 3 & 3 & 3 & - & 4273.18 & 4194 \\
    \bottomrule
    \end{tabular}
    \caption{Numbers of selected variables for distribution parameters $\mu_1$, $\sigma_1$, $\mu_2$, $\sigma_2$ and~$\rho$, negative log-likelihood values and stopping iteration~$m_{\rm{stop}}$ for classical boosting, deselection with threshold values of 0.1\% and 1\%, probing and stability selection.} \label{Tab:BirthData}
\end{table}

Table~\ref{Tab:BirthData} shows the numbers of selected variables for each distribution parameter, the predictive performance in terms of the negative log-likelihood as well as the resulting optimal $m_{\text{stop}}$. An overview of the included variables for the different approaches can be found in the supplement: Section D. The classical boosted copula model selected almost all considered variables for the mean parameters $\mu_1$ and $\mu_2$. Fewer variables were selected for the shape parameters and the dependence parameter. 
The approaches for enhanced variable selection reduced the model complexity substantially and led to fewer included variables in the final models. The deselection approach with a threshold value of 1\% deselected all variables for the mean parameter~$\mu_1$. Still, it contained variables for the other distribution parameters, more precisely interactions of the sonographic variables and the gestational age for the scale parameters. As expected, it resulted in a slightly worse negative log-likelihood compared to the classical approach.

With a smaller threshold value (0.1\%), the final model contained variables for each distribution parameter and led to a comparable predictive performance than the classical boosted model. Here the model included mostly interactions of the sonographic variables as well but also a few other variables, e.g., sex for the location and gestational age for each distribution parameter except the dependence parameter. 
Probing resulted in a similar model as the deselection approach with 0.1\%, but led to worse predictive performance. The most likely reason is the stronger shrinkage of the effect estimates due to the much smaller number of iterations.
Via stability selection no covariates were selected as stable for the dependence parameter implying conditional independence of the two responses. This resulted in a poorer predictive performance compared to the classical model.  

\subsection{Joint modelling of cholesterol phenotypes} 
We analyzed data from the UK Biobank (application number 81202), which is a large biomedical cohort study containing genetic and health information from over half a million British participants~\citep{sudlow2015ukbiobank}. 
Using the boosting algorithm for distribution copula regression, we aim to model the polygenic contribution to the individual distributions of different phenotypes, but also to estimate the dependence between these phenotypes as a function of genetic variants. We want to identify the most relevant variants and therefore apply the methods presented in Section~\ref{Sec:VarSel} to obtain sparse solutions.

The focus in the following is on three bivariate combinations of phenotypes, namely LDL (\textit{Low-Density Lipoprotein}) and ApoB (\textit{Apolipoprotein B}), LDL and cholesterol, and HDL (\textit{High-Density Lipoprotein}) and ApoA (\textit{Apolipoprotein A}). 
We considered these combinations because they have high empirical association based on an analysis of genetic blood and urine biomarkers in the UK Biobank \citep{PhenotypeUKBiobank}, suggesting potential benefits in modelling these phenotypes jointly.
All of these phenotypes are components of cholesterol metabolism. Cholesterol can be split mainly into two groups: i) LDL cholesterol, which is responsible for the transportation of cholesterol from the liver to various tissues and can be attached to specific receptors on the cell surface with the help of ApoB,  ii) HDL cholesterol, the counterpart of LDL which is accountable for the removal of excess LDL cholesterol from the body, with ApoA supporting this process~\citep{Chol_ApoA_ApoB}.

The considered dataset for each combination of phenotypes consists of $n = 20,000$ randomly sampled observations with white British ancestry. Additionally, 15,000 observations were used for validation and 20,000 observations were used to evaluate the prediction performance via the negative log-likelihood. For each phenotype, 1000 variants were selected in a pre-screening step based on the largest marginal associations between the variants and the phenotype, which were computed with the PLINK2 function~\texttt{-variant-score}~\citep{ChangPlink,Plink2}.  Variants with minor allele frequency not less than 1\% were randomly sampled with the \texttt{-thin-count} function. Missing genotypes were imputed by the reference allele using the R package \textbf{bigsnpr}~\citep{Priv2018}. After the pre-screening, the dataset contains $1,156$ variants for LDL and ApoB (844 variants were selected for both phenotypes), $1,179$ variants for LDL and cholesterol (821 common variants), and $1,249$ variants for HDL and ApoA (751 common variants).
 
For each combination of two phenotypes, the marginal distributions and copulas were chosen which minimize the predictive risk (see Table~\ref{Tab:Pheno}). All variants were considered for each distribution parameter and incorporated with linear base-learners and step length $\nu = 0.01$. Stability selection unfortunately could not be applied to these data because of the high computational cost.

\begin{table}[t!]
\resizebox{\textwidth}{!}{%
    \centering
    \begin{tabular}{llll|ccccccc}
    \toprule
     Phenotype & Marginals & Copula & Method & $\mu_1$  & $\sigma_1$ & $\mu_2$ & $\sigma_2$ & $\rho$ & $-$Log-Lik & $m_{\rm{stop}}$\\ \midrule
    LDL & log-logistic & Gaussian & classic  & 441 & 26 & 386 & 67 & 47 & 10535.16 & 4965\\
    ApoB & Gamma & & deselection 0.1\% & 121 & 2& 71 & 9 & 15 & 10749.28 & 4965\\
       & && deselection 1\% & 8 & - & 3 & - & - & 11647.92 & 4965 \\
        && & probing & - & - & - & - & - &  14792.60 & 863\\
    \midrule   
     LDL  &  log-logistic & Gumbel & classic & 286 & 89 & 266 & 100 & 44  & 31105.03 & 13975 \\ 
     Cholesterol & log-logistic & & deselection 0.1\% & 81 & 5 & 54 & 7 & 9 & 31090.53 & 13975 \\
     & & & deselection 1\% & 12 & - & 6 & - &  - & 31817.39 & 13975 \\
     & & & probing & 45 & - & 15 & 3 & - & 32834.92 & 5314 \\
      \midrule
       HDL  & log-normal & Gaussian & classic & 171 & 40 & 197 & 69 & 28 & 79820.43 & 1954 \\  
       ApoA & log-normal & & deselection 0.1\% & 81 & 15 & 83 & 24 & 9 & 79868.42 & 1954\\
            & & &deselection 1\% & 8 & - & 9 & - & - & 80337.71 & 1954 \\
            & & &probing & 113 & 20 & 185 & 36 & 6 &  80074.35 & 905 \\
         \bottomrule
    \end{tabular}}
    \caption{Number of selected variants for distribution parameters~$\mu_1$,  $\sigma_1$, $\mu_2$, $\sigma_2$ and~$\rho$, negative log-likelihood values and stopping iteration~$m_{\rm{stop}}$ for the classical boosted model, the deselection approach with threshold values of 0.1\% and 1\%, and probing for the different combinations of phenotypes.}\label{Tab:Pheno}
\end{table}

\begin{figure}[t!]\centering
    \includegraphics[width=\textwidth]{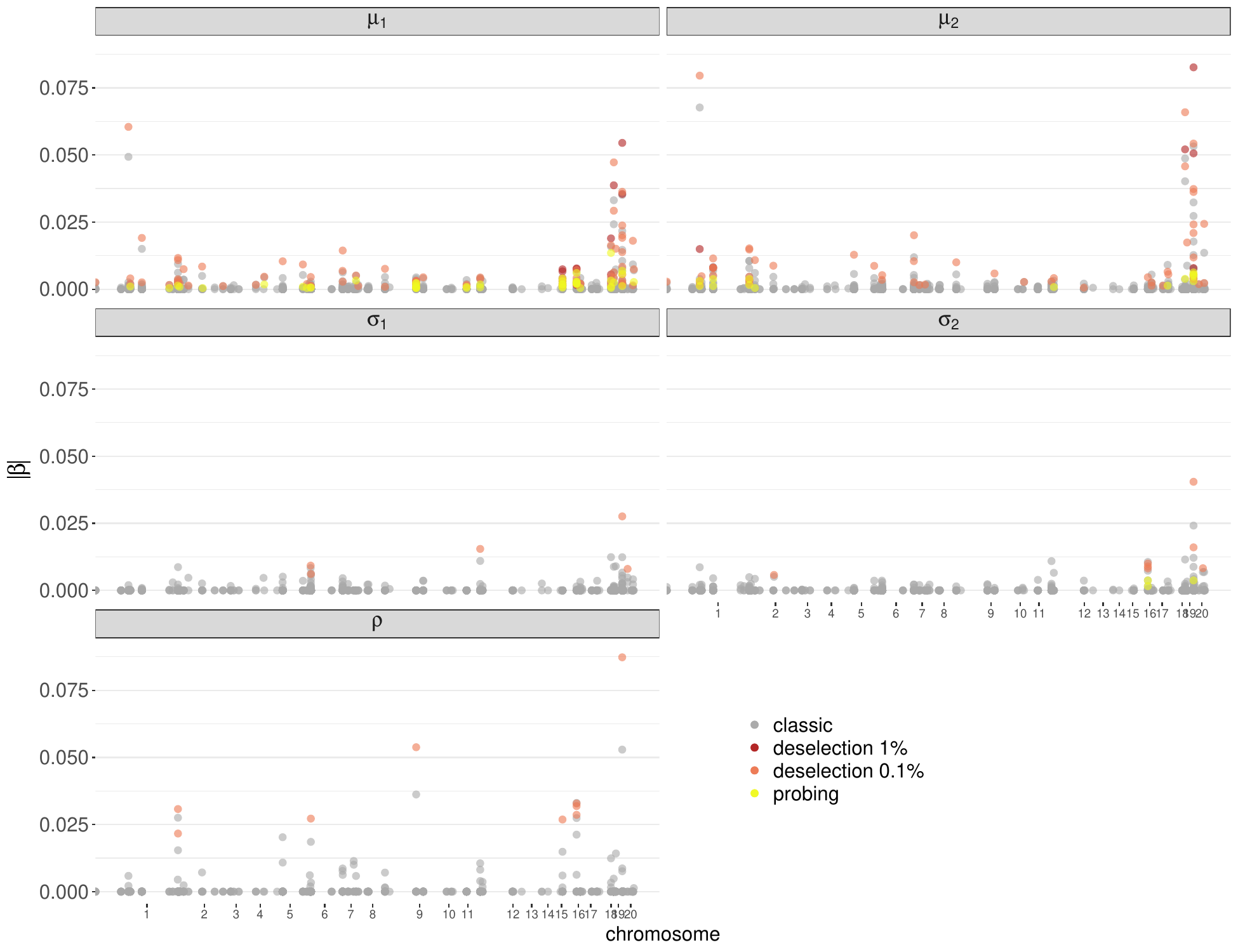}
    \caption{Manhattan-type plots (chromosomes on x-axis) for the absolute coefficients of boosted copula regression for the joint analysis of LDL and cholesterol. 
    }\label{Fig:Variants_LDL_Chol}
\end{figure}

Table~\ref{Tab:Pheno} shows the results for the joint analysis of the different combinations of phenotypes. Furthermore, Figure~\ref{Fig:Variants_LDL_Chol} displays Manhattan-type plots for the phenotype combination LDL and cholesterol for every distribution parameter of the copula model. For each combination of phenotypes, the classical boosting approach selected several variants for each distribution parameter. Most genetic variants were selected for the location parameters. Each model included variants for the dependence parameter, indicating that different variants affect the associations between phenotypes and the potential benefit of modelling these phenotypes together. 
Considering the total number of selected variants, a relatively high number of the pre-filtered variants were included in the classical boosting model. In particular, for LDL and ApoB, almost half of all variants were included for the mean parameters. Despite the intrinsic variable selection of the boosting algorithm, we still obtain large models with a potentially difficult interpretation. Therefore, we aim to reduce the model complexity by enhancing variable selection.

With the deselection approach, the model complexity could be drastically reduced. When considering a threshold value of 1\%, for all phenotype combinations only variants for the location parameters remained after deselection, which resulted in two univariate models. One can argue that this threshold value may be too strong for the data situation as there are several variants with only a small to medium effect (see for example Figure~\ref{Fig:Variants_LDL_Chol}) and therefore a minor impact on risk reduction. Also, owing to the pre-filtering, all variants in our analysis have some association with one of the outcomes, making it harder for single variants to pass the relative threshold.
Using a smaller threshold value (0.1\%) also led to sparser models, but for each distribution parameter several variants remained in the final model. The negative log-likelihood indicated a comparable predictive performance, whereby even a slight improvement in the performance for the phenotype combination LDL and cholesterol could be observed.

Probing also resulted in sparser models for each combination. In fact, for the phenotypes LDL and ApoB, no variants were included in the resulting model: in the first iterations, only the intercept was updated and stopping after 863 boosting iterations (when the first probe was selected) resulted in an intercept model, which led to a considerably worse predictive performance. For the other phenotypes, several variants were included after stopping when the first probe was selected. However, due to the smaller number of boosting iterations, the effect estimates were more shrunken (see Figure~\ref{Fig:Variants_LDL_Chol} for LDL and cholesterol) and therefore the predictive performance deteriorated in comparison to the classical boosted model but also to the deselection approach, particularly for a threshold value of 0.1\%.

\section{Discussion and conclusion}

To reduce model complexity and to enhance variable selection for boosting multivariate distributional copula regression, we have integrated  probing \citep{Thomas2017}, stability selection \citep{Meinshausen2010}, and also the recent deselection approach \citep{Deselection} in the boosting framework for this model class. This combination of classical boosting with all three approaches leads to considerably sparser models, thereby improving the interpretability of the obtained prediction models, which is desirable in practice~\citep{wyatt1995commentary,Markowetz2024}.

Regarding the specific approaches, the results of stability selection show similarities to the ones from deselection, even though the initial goals of the two methods differ. All three approaches perform better when the true model is sparse, whereby deselection can still lead to reasonable results when many variables are informative. The probing approach is the most favorable regarding computational runtime, but typically stops the algorithm also very early, leading often to underfitting the data and reduced predictive performance. As also observed in our first application on the weight and length of newborns, stability selection and deselection are more often able to maintain the predictive performance with smaller models. However, only deselection is also scalable to large, high-dimensional data as in our genetic application.

Our results additionally suggest that deselection not only yields much sparser models but can even lead to simpler univariate regression models in comparison to the classical boosted copula model in situations where the association parameter is close to zero. The proposed methods for enhanced variable selection could hence also represent tools for data driven model choice \citep{mayr2023linear}. The prediction performance typically does not improve after deselection but can lead to comparable accuracy as the classical boosting model with fewer predictors. The procedure is controlled via a threshold value~$\tau$, which represents the minimum amount of total risk reduction that should be attributed to a corresponding base-learner to avoid deselection. This can be interpreted as a threshold value for the importance of the particular predictor variable. Depending on the data situation, different thresholds may be appropriate; however, tuning is not straightforward because the true number of informative variables is not known in practice and the best model regarding predictive risk is naturally the one without any deselection. Further research is warranted on how to specify the threshold~$\tau$ in this context.  


Besides the practical advantages of the proposed tools, the natural limitation of all boosting algorithms applies: Due to early stopping and therefore shrinkage of the effect estimates, providing standard errors of the resulting coefficients is not an easy task as there are no closed formulas. To overcome this, one could apply permutation tests to carry out significance testing and provide $p$-values~\citep{Mayr2017permutation}, but this would drastically increase the computational cost. 

In conclusion, while statistical models should be as complex as needed to be able to capture the underlying nature of the data-generating process, they should also remain as simple as possible to facilitate interpretation. To navigate this conceptual trade-off, we have proposed three competing approaches to simplify distributional copula regression models by reducing the model complexity and to enhance the variable selection properties of statistical boosting without considerably reducing the prediction accuracy of the resulting models.




\subsubsection*{Availability of data and materials}
The code used for the simulations and the biomedical applications is available at GitHub~\url{https://github.com/AnnikaStr/ComplRedBoostCop}. The fetal ultrasound data is not publicly available. The genomic cohort data is available upon request from the UK Biobank at \url{https://www.ukbiobank.ac.uk/}.

\subsubsection*{Funding}
Open Access funding enabled and organized by Projekt DEAL. The work on this article was supported by the Deutsche Forschungsgemeinschaft (DFG, grant number 428239776, KL3037/2-1, MA7304/1-1).

\bibliography{EnhancedVariableSelection_Stroemer}
\bibliographystyle{apalike}

\newpage
\clearpage

\appendix

\setcounter{figure}{0}

\makeatletter 
\renewcommand{\thefigure}{A\@arabic\c@figure}
\makeatother

\makeatletter 
\renewcommand{\thetable}{A\@arabic\c@table}
\makeatother

\section*{\centering \LARGE Supplementary Materials}
\addcontentsline{toc}{section}{\protect\numberline{}Supplementary Materials}

\section{Example}
Toy example for the method section: We consider a low-dimensional setting with $p=20$ variables and the predictors were as follows:
\begin{align*}
    \mu_1 = -x_1 + 0.5 x_3 & \qquad \mu_2 = -0.5 - 0.7 x_1 - 0.3 x_2 \\
    \sigma_1 = -0.7 - 0.7 x_3 & \qquad \sigma_2= 2 + 0.5 x_2\\
    \rho  &= 1 + 1x_4,
\end{align*}
where the covariates were sampled independently of a uniform distribution on $(-1,1)$ with four informative variables $x_1, \dots, x_4$ and $n = 1000$ number of observations. The number of boosting iterations was optimized on a validation data set with $1500$ observations. The step length was set to $\nu=0.01$ and all covariates were included with linear base-learners. 
For stability selection, the number of variables to be included in the model is set as $q =20$ and the per-family error-rate was chosen with $\mathrm{PFER} = 5$. 

\begin{figure}[h]
    \centering
    \includegraphics[width=1\linewidth]{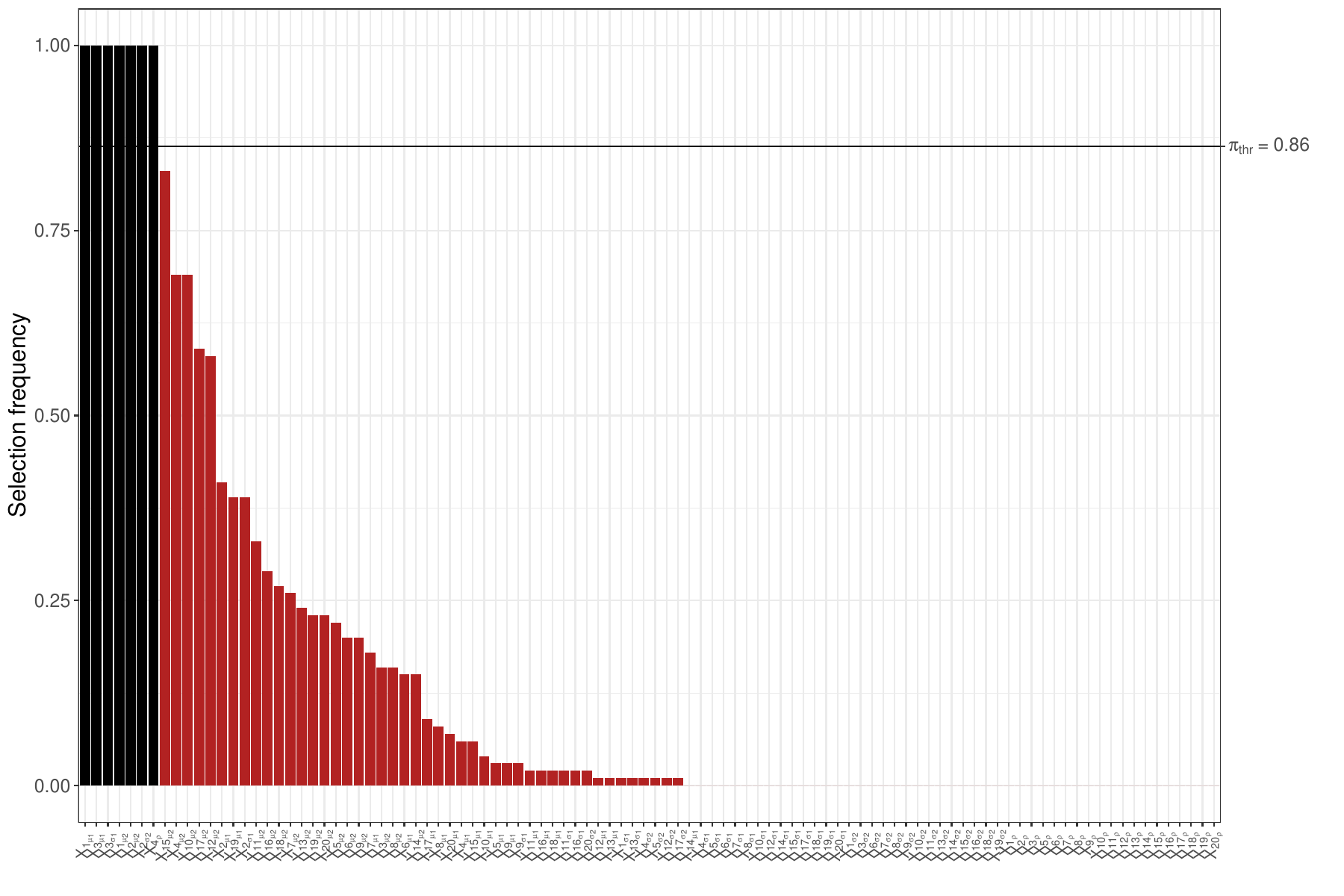}
    \caption{Selection frequencies of the covariates over the $B$ subsets for the simulated data example for stability selection. }
\end{figure}

\begin{figure}
    \centering
    \includegraphics[width=\linewidth]{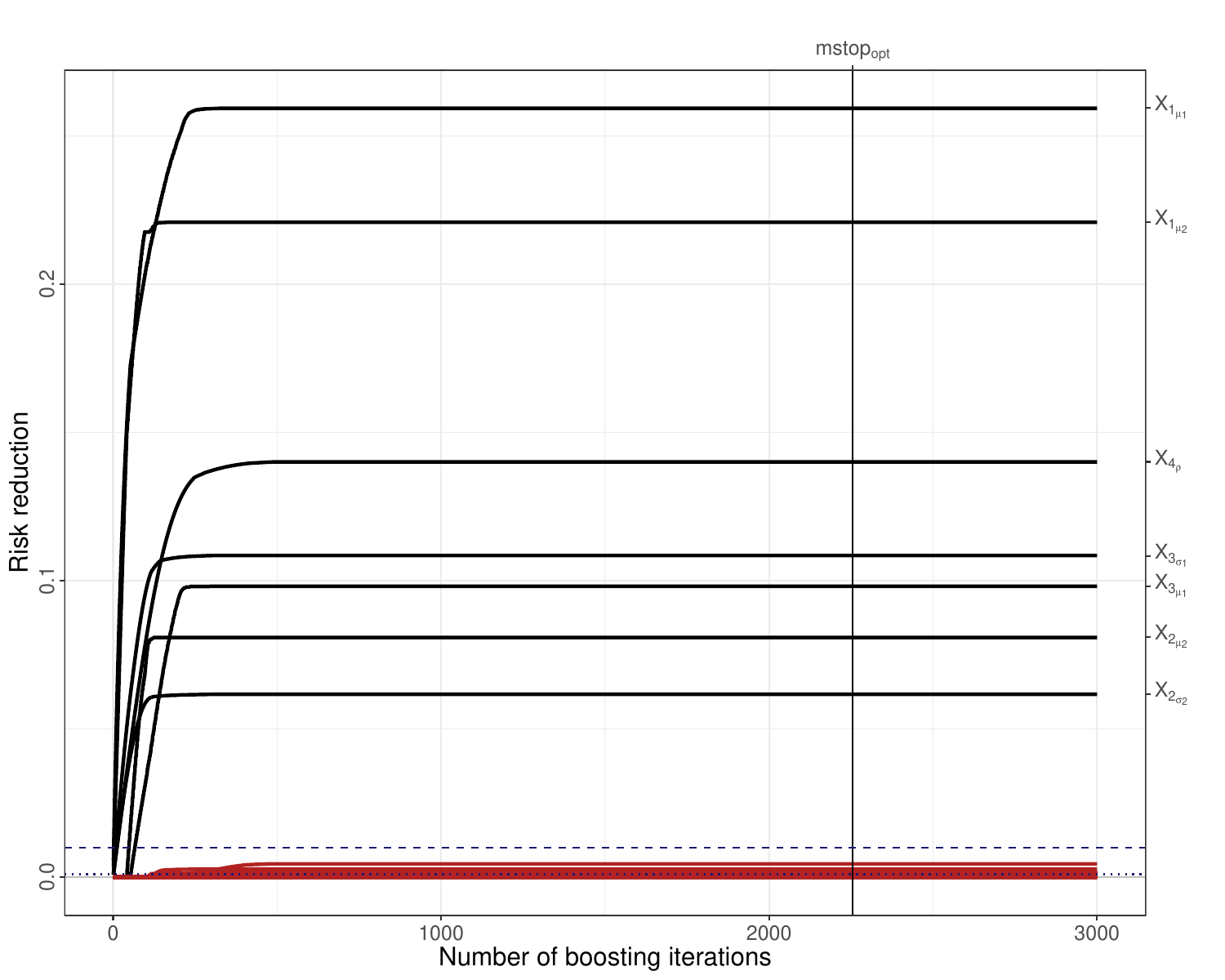}
    \caption{Risk reduction for the toy example. The red coefficient path corresponds to the non-informative variables resulting from a threshold value of 1\%.}
\end{figure}


\section{Simulations}

\subsection{Scenario A}
Following the simulation design of Hans et al., we carried out a simulation study for boosted copula regression with the Gaussian, Clayton, and Gumbel copula in combination with the Log-Normal and Log-Logistic marginal distributions. For every setting, a low-dimensional case ($p<n$) with $p = 20$ variables and a high-dimensional case ($p>n$) with $p = 1000$ variables were considered. The data were generated from the following predictors
\begin{align*}
    \mu_1 = -0.75x_1 + 0.5\cos(\pi x_3) & \qquad \mu_2 = 0.5- 0.7 x_1 - 0.02 \exp (2(x_2 +1)) \\
    \sigma_1 = -0.7 + 0.5 \sin (\pi x_3) & \qquad \sigma_2= 2 + 0.5 x_2\\
    \rho  = - 0.8 & + 1.5 \log (4.5 - 1.7\sin (\pi x_4)),
\end{align*}
where the covariates were sampled independently of a uniform distribution on $(-1,1)$ with four informative variables $x_1, \dots, x_4$ and $n = 1000$ number of observations. 
As base-learners, we applied cubic P-splines with 20 equidistant knots and a second-order difference penalty. 

\begin{table}[ht]
\centering
\resizebox{\textwidth}{!}{%
\begin{tabular}{rlcccccccccccccccc}
  \toprule
  & & \multicolumn{5}{c}{True Positives}  & \multicolumn{5}{c}{False Positives} & & & &\\ 
 & method & $\mu_1$ & $\sigma_1$ & $\mu_2$ & $\sigma_2$ & $\rho$ & $\mu_1$ & $\sigma_1$ & $\mu_2$ & $\sigma_2$ & $\rho$ & -Log-Lik & ES & time  \\ 
  \midrule
 \multirow{5}{*}{\rotatebox[origin=c]{90}{Clayton}} & classic & 2 (0) & 1 (0) & 2 (0) & 1 (0) & 1 (0) & 10.68 (2.63) & 5.94 (2.18) & 17.28 (1.03) & 4.24 (1.69) & 0.39 (0.6) & 237.11 (54.79) & 0.51 (0.02) & 12.32\\ 
   & deselection $\tau = 0.01$
   & 2 (0) & 1 (0) & 2 (0) & 1 (0) & 1 (0) & 0 (0) & 0 (0) & 0 (0) & 2 (0) & 0 (0) &  196.24 (55.89) & 0.51 (0.02) & 13.59\\ 
   & deselection $\tau=0.001$ & 2 (0) & 1 (0) & 2 (0) & 1 (0) & 1 (0) & 0.02 (0.14) & 0.18 (0.39) & 1.03 (0.17) & 2.03 (0.17) & 0.12 (0.33) &  199.46 (56.40) & 0.51 (0.02) & 13.63 \\ 
   & probing & 2 (0) & 1 (0) & 2 (0) & 1 (0) & 0.42 (0.50) & 0 (0) & 0.04 (0.20) & 2.12 (1.40) & 2 (0) & 0 (0) &  586.19 (94.71) & 0.53 (0.02) & 2.48 \\ 
   & stabsel\_q20\_pfer5 & 2 (0) & 1 (0) & 2 (0) & 1 (0) & 0.88 (0.33) & 0 (0) & 0 (0) & 1.30 (0.58) & 2 (0) & 0 (0) &  192.77 (62.95) & 0.51 (0.02) &  146.20 \\ \midrule

   \multirow{5}{*}{\rotatebox[origin=c]{90}{Gauss}}  & classic & 2 (0) & 1 (0) & 2 (0) & 1 (0) & 1 (0) & 6.07 (2.56) & 3.67 (1.87) & 17.49 (0.66) & 3.45 (1.51) & 1.45 (0.58) &  651.35 (60.47) & 0.50 (0.03) & 10.10\\ 
   & deselection $\tau =0.01$ & 2 (0) & 1 (0) & 2 (0) & 1 (0) & 1 (0) & 0 (0) & 0 (0) & 0 (0) & 1 (0) & 1.08 (0.34) & 635.86 (60.16) & 0.50 (0.02) & 11.09  \\ 
   & deselection $\tau = 0.001$ & 2 (0) & 1 (0) & 2 (0) & 1 (0) & 1 (0) & 0.01 (0.1) & 0.24 (0.47) & 1.08 (0.27) & 1.12 (0.33) & 1.36 (0.5) & 641.07 (60.03) & 0.50 (0.02) & 11.18 \\ 
   & probing & 2 (0) & 0.99 (0.10) & 2 (0) & 1 (0) & 0.99 (0.10) & 0 (0) & 0.05 (0.22) & 1.86 (1.18) & 1 (0) & 1.36 (0.52) & 823.98 (115.06) & 0.52 (0.03) & 3.18 \\ 
   & stabsel\_q20\_pfer5 & 2 (0) & 1 (0) & 2 (0) & 1 (0) & 1 (0) & 0 (0) & 0 (0) & 1.43 (0.62) & 1 (0) & 0.24 (0.43) & 632.04 (59.92) & 0.50 (0.02) &  115.93 \\ \midrule

    \multirow{5}{*}{\rotatebox[origin=c]{90}{Gumbel}}  & classic & 2 (0) & 1 (0) & 2 (0) & 1 (0) & 1 (0) & 14.84 (2.02) & 8.49 (2.25) & 11.12 (2.52) & 6.16 (1.87) & 0.14 (0.35) & 314.04 (80.24) & 0.50 (0.03) & 26.50 \\ 
   & deselection $\tau =0.01$ & 2 (0) & 1 (0) & 2 (0) & 1 (0) & 1 (0) & 0 (0) & 0 (0) & 0 (0) & 1.82 (0.39) & 0 (0) & 275.73 (77.79) & 0.50 (0.03) & 29.07\\ 
   & deselection $\tau = 0.001$ & 2 (0) & 1 (0) & 2 (0) & 1 (0) & 1 (0) & 0.01 (0.10) & 0.34 (0.54) & 1 (0) & 2.02 (0.20) & 0.03 (0.17) & 282.46 (80.29) & 0.50 (0.03) & 29.06 \\ 
   & probing & 2 (0) & 1 (0) & 2 (0) & 1 (0) & 0.81 (0.39) & 0 (0) & 0.12 (0.33) & 1.87 (1.40) & 2 (0.14) & 0 (0) & 491.91 (82.91) & 0.52 (0.03) & 7.19 \\ 
  & stabsel\_q20\_pfer5 & 2 (0) & 1 (0) & 2 (0) & 1 (0) & 1 (0) & 0 (0) & 0.02 (0.14) & 1.22 (0.54) & 1.88 (0.33) & 0 (0) & 280.14 (79.88) & 0.50 (0.02) & 263.84 \\ 
   \bottomrule
\end{tabular}}
\caption{Low-dimensional setting: Results of the classical boosted model, the deselection approach, stability selection and probing for the Clayton, Gaussian and Gumbel Copula in terms of mean (sd) of the true positives, false positives and the predictive performance in terms of negative log-likelihood and energy score (ES). The computational time corresponds to one simulation run.  }
\end{table}

\begin{table}[ht]
\centering
\resizebox{\textwidth}{!}{%
\begin{tabular}{rlccccccccccccccc}
  \toprule
  & & \multicolumn{5}{c}{True Positives}  & \multicolumn{5}{c}{False Positives} & & & &\\ 
 & method & $\mu_1$ & $\sigma_1$ & $\mu_2$ & $\sigma_2$ & $\rho$ & $\mu_1$ & $\sigma_1$ & $\mu_2$ & $\sigma_2$ & $\rho$ &  -Log-Lik & ES & time \\ 
  \midrule

\multirow{4}{*}{\rotatebox[origin=c]{90}{Clayton}} & classic & 2 (0) & 1 (0) & 2 (0) & 1 (0) & 1 (0) & 0.13 (0.39) & 0.79 (1.02) & 104.56 (12.37) & 2.92 (1.02) & 0.12 (0.43) & 447.93 (60.37) & 0.52 (0.03) &  103.79\\ 
   & deselection $\tau  = 0.01$ & 2 (0) & 1 (0) & 2 (0) & 1 (0) & 1 (0) & 0 (0) & 0 (0) & 0 (0) & 2 (0) & 0.04 (0.20) & 315.45 (56) & 0.52 (0.02) &  104.85\\ 
   & deselection $\tau = 0.001$ & 2 (0) & 1 (0) & 2 (0) & 1 (0) & 1 (0) & 0 (0) & 0.05 (0.22) & 0.93 (0.26) & 2.13 (0.37) & 0.04 (0.20) &  319.77 (56.72) & 0.52 (0.02) & 104.82 \\ 
   & probing & 2 (0) & 1 (0) & 2 (0) & 1 (0) & 0.05 (0.22) & 0 (0) & 0.12 (0.36) & 16.42 (15.64) & 2.12 (0.33) & 0 (0) & 684.64 (72.49) & 0.53 (0.03) & 29.12 \\ \midrule
  
  \multirow{4}{*}{\rotatebox[origin=c]{90}{Gauss}}  & classic & 2 (0) & 1 (0) & 2 (0) & 1 (0) & 1 (0) & 0 (0) & 0.02 (0.14) & 64.58 (11.77) & 1.08 (0.27) & 1.71 (0.84) &  816.5 (56.08) & 0.52 (0.02) & 102.85 \\ 
   & deselection $\tau =0.01$ & 2 (0) & 1 (0) & 2 (0) & 1 (0) & 1 (0) & 0 (0) & 0 (0) & 0 (0) & 1 (0) & 1.18 (0.48) & 737.16 (57.09) & 0.51 (0.02) &  103.64 \\ 
    & deselection $\tau = 0.001$ & 2 (0) & 1 (0) & 2 (0) & 1 (0) & 1 (0) & 0 (0) & 0 (0) & 1 (0.20) & 1.07 (0.26) & 1.63 (0.75) & 746.96 (57.80) & 0.52 (0.02) & 103.84 \\ 
   & probing & 1.99 (0.10) & 0.93 (0.26) & 2 (0) & 1 (0) & 0.93 (0.26) & 0 (0) & 0.01 (0.10) & 15.57 (15.70) & 1.08 (0.27) & 1.66 (0.88) & 895.51 (185.05) & 0.54 (0.04) & 28.52 \\ \midrule

   \multirow{4}{*}{\rotatebox[origin=c]{90}{Gumbel}} & classic & 2 (0) & 1 (0) & 2 (0) & 1 (0) & 1 (0) & 7.04 (7.43) & 1.21 (1.58) & 69.97 (9.21) & 4.26 (1.96) & 0.13 (0.42) & 500.87 (80.6) & 0.52 (0.02) & 118.55 \\ 
   & deselection $\tau = 0.01$ & 2 (0) & 1 (0) & 2 (0) & 1 (0) & 1 (0) & 0 (0) & 0 (0) & 0 (0) & 1.84 (0.37) & 0 (0) & 333.88 (80.18) & 0.50 (0.02) &  120.70\\ 
    & deselection $\tau = 0.001$ & 2 (0) & 1 (0) & 2 (0) & 1 (0) & 1 (0) & 0 (0) & 0.05 (0.22) & 1 (0) & 2.14 (0.43) & 0 (0) & 344.75 (80.92) & 0.50 (0.02) & 120.65 \\ 
   & probing & 2 (0) & 1 (0) & 2 (0) & 1 (0) & 0.31 (0.46) & 0 (0) & 0.19 (0.61) & 10.25 (10.84) & 2.14 (0.51) & 0 (0) & 569.17 (81.61) & 0.52 (0.03) & 51.61\\ 
   \bottomrule
\end{tabular}}
\caption{High-dimensional setting: Results of the classical boosted model, the deselection approach with $\tau =0.01$ and probing for the Clayton, Gaussian and Gumbel Copula in terms of mean (sd) of the true positives, false positives and the predictive performance in terms of negative log-likelihood and energy score (ES). The computational time corresponds to one simulation run. }
\end{table}

\newpage
\subsection{Scenario B} 

The data were generated from the following predictors
\begin{align*}
    \mu_1 = -0.75x_1 + 0.5\cos(\pi x_3)  &\qquad \mu_2 = 0.5- 0.7 x_1 - 0.02 \exp (2(x_2 +1)) \\
    \sigma_1 = -0.7  &\qquad  \sigma_2 =  2 + 0.5 x_2\\
    \rho  = - 0.8  &+ 1.5 \log (4.5 - 1.7\sin (\pi x_4)),
\end{align*}
where the covariates were sampled independently of a uniform distribution on $(-1,1)$ with four informative variables $x_1, \dots, x_4$ and $n = 1000$ number of observations. 
As base-learners, we applied cubic P-splines with 20 equidistant knots and a second-order difference penalty.

\begin{table}[ht]
\centering
\resizebox{\textwidth}{!}{%
\begin{tabular}{rlccccccccccccccc}
  \toprule
  & & \multicolumn{5}{c}{True Positives}  & \multicolumn{5}{c}{False Positives} & & & &\\ 
 & method & $\mu_1$ & $\sigma_1$ & $\mu_2$ & $\sigma_2$ & $\rho$ & $\mu_1$ & $\sigma_1$ & $\mu_2$ & $\sigma_2$ & $\rho$ & -Log-Lik & ES & time  \\ 
  \midrule
\multirow{5}{*}{\rotatebox[origin=c]{90}{Clayton}} & classic & 2 (0) & 0 (0) & 2 (0) & 1 (0) & 1 (0) & 7.66 (2.65) & 12.25 (1.86) & 17.46 (0.7) & 4.27 (1.67) & 0.36 (0.61) &  228.28 (54.87) & 0.47 (0.02) &  12.80 \\ 
   & deselection $\tau = 0.01$ & 2 (0) & 0 (0) & 2 (0) & 1 (0) & 1 (0) & 0 (0) & 1.08 (0.27) & 0 (0) & 2 (0) & 0 (0) & 190.18 (55.30) & 0.46 (0.02) &  13.97 \\ 
   & deselection $\tau = 0.001$ & 2 (0) & 0 (0) & 2 (0) & 1 (0) & 1 (0) & 0.04 (0.2) & 5.84 (1.35) & 1.01 (0.17) & 2.01 (0.1) & 0.12 (0.33) &  197.72 (56.04) & 0.47 (0.02) & 14.00 \\ 
   & probing & 2 (0) & 0 (0) & 2 (0) & 1 (0) & 0 (0) & 0 (0) & 2.74 (1.8) & 0.98 (0.25) & 2 (0) & 0 (0) & 883.1 (133.25) & 0.51 (0.03) &  2.44 \\ 
   & stabsel\_q20\_pfer5 & 2 (0) & 0 (0) & 2 (0) & 1 (0) & 0 (0) & 0 (0) & 0.84 (0.71) & 0.8 (0.4) & 1.96 (0.2) & 0 (0) & 246.46 (57.88) & 0.46 (0.02) &  104.89 \\ \midrule

 \multirow{5}{*}{\rotatebox[origin=c]{90}{Gauss}}& classic & 2 (0) & 0 (0) & 2 (0) & 1 (0) & 1 (0) & 3.66 (2.2) & 10.38 (1.87) & 17.59 (0.71) & 3.9 (1.56) & 1.02 (0.32) & 669.73 (56.19) & 0.46 (0.02) & 10.52 \\ 
   & deselection $\tau = 0.01$ & 2 (0) & 0 (0) & 2 (0) & 1 (0) & 1 (0) & 0 (0) & 1.14 (0.35) & 0 (0) & 1 (0) & 0.9 (0.33) & 656.18 (55.48) & 0.46 (0.02) &  11.41 \\ 
   & deselection $\tau = 0.001$ & 2 (0) & 0 (0) & 2 (0) & 1 (0) & 1 (0) & 0.02 (0.14) & 4.49 (1.02) & 1.13 (0.34) & 1.06 (0.28) & 0.98 (0.28) & 660.15 (56.04) & 0.46 (0.02) &  11.43 \\ 
   & probing & 2 (0) & 0 (0) & 2 (0) & 1 (0) & 0.92 (0.27) & 0 (0) & 2.37 (1.10) & 1.25 (0.77) & 1 (0) & 1 (0.32) &  907.59 (140.54) & 0.49 (0.03) & 7.47 \\ 
   & stabsel\_q20\_pfer5 & 2 (0) & 0 (0) & 2 (0) & 1 (0) & 1 (0) & 0 (0) & 1.14 (0.65) & 1.08 (0.27) & 1 (0) & 0.19 (0.39) & 639.07 (56.38) & 0.45 (0.02) & 120.66 \\ \midrule

   \multirow{5}{*}{\rotatebox[origin=c]{90}{Gumbel}} & classic & 2 (0) & 0 (0) & 2 (0) & 1 (0) & 1 (0) & 12.67 (2.96) & 13.29 (1.83) & 10.47 (2.95) & 6.21 (2.04) & 0.12 (0.41) & 304.33 (79.99) & 0.45 (0.02) & 25.91  \\ 
   & deselection $\tau = 0.01$ & 2 (0) & 0 (0) & 2 (0) & 1 (0) & 1 (0) & 0 (0) & 1.06 (0.24) & 0 (0) & 1 (0) & 0 (0) & 260.02 (75.64) & 0.45 (0.02) &  25.33\\ 
   & deselection $\tau =0.001$ & 2 (0) & 0 (0) & 2 (0) & 1 (0) & 1 (0) & 0.01 (0.10) & 5.08 (1.17) & 1 (0) & 1.09 (0.29) & 0.01 (0.10) & 273 (79.43) & 0.45 (0.02) &  25.40\\ 
   & probing & 2 (0) & 0 (0) & 2 (0) & 1 (0) & 0.06 (0.24) & 0.02 (0.14) & 3.06 (1.26) & 1.02 (0.14) & 1.01 (0.10) & 0 (0) & 592.59 (87.35) & 0.49 (0.02) &  4.38 \\ 
   & stabsel\_q20\_pfer5 & 2 (0) & 0 (0) & 2 (0) & 1 (0) & 0.95 (0.22) & 0 (0) & 1.63 (0.54) & 1 (0) & 1 (0) & 0 (0) & 278.23 (81.69) & 0.45 (0.02) & 205.80\\ 
   \bottomrule
\end{tabular}}
\caption{Low-dimensional setting: Results of the classical boosted model, the deselection approach with $\tau =0.01$, stability selection and probing for the Clayton, Gaussian and Gumbel Copula in terms of mean (sd) the true positives, false positives and the predictive performance in terms of negative log-likelihood and energy score (ES). The computational time correpesonds to one simulation run. }
\end{table}

\begin{table}[ht]
\centering
\resizebox{\textwidth}{!}{%
\begin{tabular}{rlccccccccccccccc}
  \toprule
  & & \multicolumn{5}{c}{True Positives}  & \multicolumn{5}{c}{False Positives} & & & &\\ 
 & method & $\mu_1$ & $\sigma_1$ & $\mu_2$ & $\sigma_2$ & $\rho$ & $\mu_1$ & $\sigma_1$ & $\mu_2$ & $\sigma_2$ & $\rho$ & -Log-Lik & ES & time \\ 
  \midrule
\multirow{4}{*}{\rotatebox[origin=c]{90}{Clayton}} &  classic & 2 (0) & 0 (0) & 2 (0) & 1 (0) & 1 (0) & 0.03 (0.17) & 20.08 (3.48) & 92.93 (13.04) & 2.65 (0.81) & 0.14 (0.38) & 407.19 (57.2) & 0.48 (0.02) & 108.09  \\ 
   & deselection $\tau = 0.01$ & 2 (0) & 0 (0) & 2 (0) & 1 (0) & 1 (0) & 0 (0) & 1 (0) & 0 (0) & 1.99 (0.1) & 0.01 (0.1) & 298.21 (56.98) & 0.47 (0.02) & 109.08 \\ 
  & deselection $\tau = 0.001$ & 2 (0) & 0 (0) & 2 (0) & 1 (0) & 1 (0) & 0 (0) & 9.91 (1.69) & 0.8 (0.4) & 2.05 (0.22) & 0.06 (0.24) & 302.97 (58.05) & 0.47 (0.02) & 109.09 \\ 
   & probing & 2 (0) & 0 (0) & 2 (0) & 1 (0) & 0.01 (0.1) & 0 (0) & 9.66 (6.15) & 4.13 (8.02) & 2.03 (0.17) & 0 (0) &  838.14 (162.85) & 0.51 (0.03) & 29.89\\ \midrule

 \multirow{4}{*}{\rotatebox[origin=c]{90}{Gauss}} & classic & 2 (0) & 0 (0) & 2 (0) & 1 (0) & 1 (0) & 0 (0) & 7.2 (2.62) & 57.94 (9.78) & 1.05 (0.22) & 1.45 (0.77) &  829.91 (52.66) & 0.48 (0.02) & 104.04 \\ 
   & deselection  $\tau = 0.01$ & 2 (0) & 0 (0) & 2 (0) & 1 (0) & 1 (0) & 0 (0) & 1.02 (0.14) & 0 (0) & 1 (0) & 1.02 (0.38) & 759.18 (52.47) & 0.47 (0.02) & 104.80 \\ 
    & deselection $\tau = 0.001$ & 2 (0) & 0 (0) & 2 (0) & 1 (0) & 1 (0) & 0 (0) & 3.99 (1.57) & 1.01 (0.1) & 1.04 (0.2) & 1.37 (0.66) & 770.63 (53.49) & 0.47 (0.02) &  104.83\\ 
   & probing & 2 (0) & 0 (0) & 2 (0) & 1 (0) & 0.89 (0.31) & 0 (0) & 4.31 (2.95) & 9.63 (12.17) & 1.06 (0.24) & 1.38 (0.83) & 889.08 (245.99) & 0.5 (0.03) & 29.51\\ \midrule

   \multirow{4}{*}{\rotatebox[origin=c]{90}{Gumbel}} & classic & 2 (0) & 0 (0) & 2 (0) & 1 (0) & 1 (0) & 4.19 (5.82) & 17.51 (3.67) & 53.8 (7.1) & 3.2 (1.99) & 0.02 (0.14) & 418.98 (72.49) & 0.47 (0.02) & 116.68 \\ 
   & deselection $\tau = 0.01$ & 2 (0) & 0 (0) & 2 (0) & 1 (0) & 1 (0) & 0 (0) & 1 (0) & 0 (0) & 1.01 (0.1) & 0 (0) & 309.19 (72.94) & 0.46 (0.02) & 118.65 \\ 
    & deselection $\tau = 0.001$ & 2 (0) & 0 (0) & 2 (0) & 1 (0) & 1 (0) & 0 (0) & 7.52 (1.37) & 1 (0) & 1.04 (0.2) & 0 (0) & 315.47 (73.27) & 0.46 (0.02) &  118.59\\ 
   & probing & 2 (0) & 0 (0) & 2 (0) & 1 (0) & 0.11 (0.31) & 0 (0) & 8.32 (5.18) & 3.09 (5.85) & 1.13 (0.6) & 0 (0) & 597.08 (111.56) & 0.49 (0.02) & 32.40 \\ 
   \bottomrule
\end{tabular}}
\caption{High-dimensional setting: Results of the classical boosted model, the deselection approach with $\tau =0.01$, stability selection and probing for the Clayton, Gaussian and Gumbel Copula in terms of mean (sd) of the true positives, false positives and the predictive performance in terms of negative log-likelihood and energy score (ES). The computational time corresponds to one simulation run. }
\end{table}

\subsection{Scenario C}
The data were generated from the following predictors
\begin{scriptsize}
\begin{align*}
    \mu_1 &= 0.5x_1 +  x_2 -0.5x_3 - z_4 + x_5 + 0.5x_6- 0.5x_7 +x_8 + 0.5x_9 + x_{10} \\ 
    \mu_2 &=  -x_{21} + 0.5 x_{22} + 0.5 x_{23} - x_{24}-0.5 x_{25} - 0.5 x_{26} -x_{27} - 0.5x_{28} + 0.5x_{29}- x_{30} \\
    \sigma_1 &= 0.5 x_{11} + 0.25x_{12} + 0.25x_{13} + 0.25x_{14}+ 0.5x_{15} + 0.25x_{16}- 0.25 x_{17} + 0.5 x_{18} - 0.25 x_{19} + 0.5 x_{20}\\
    \sigma_2 & = -0.25x_{31} + 0.25 x_{32} + 0.25x_{33} +0.25 x_{34} - 0.5x_{35} -0.25x_{36} +0.25x_{37} + 0.25 x_{38} - 0.5x_{39} - 0.5x_{40} \\
    \rho  &= - 0.5x_{41} - x_{42} + 0.5 x_{43} + 0.5 x_{44}-0.5x_{45} - x_{46} + 0.5 x_{47} + 0.5 x_{48} + x_{49} - x_{50},
\end{align*}
\end{scriptsize}
where the covariates were sampled independently of a uniform distribution on $(-1,1)$ with ten informative variables for each parameter and $n = 1000$ number of observations. 
The base-learners correspond to simple linear models.

\begin{table}[ht]
\centering
\resizebox{\textwidth}{!}{%
\begin{tabular}{rlcccccccccccccccc}
  \toprule
  & & \multicolumn{5}{c}{True Positives}  & \multicolumn{5}{c}{False Positives} & & & &\\ 
 & method & $\mu_1$ & $\sigma_1$ & $\mu_2$ & $\sigma_2$ & $\rho$ & $\mu_1$ & $\sigma_1$ & $\mu_2$ & $\sigma_2$ & $\rho$ & -Log-Lik & ES & time  \\ 
  \midrule
\multirow{5}{*}{\rotatebox[origin=c]{90}{Clayton}} & classic & 10 (0) & 10 (0) & 10 (0) & 10 (0) & 4.6 (1.5) & 39.07 (8.48) & 40.97 (7.17) & 35.56 (7.94) & 40.43 (5.92) & 1.01 (0.1)   & 2705.83 (55.2) & 1.11 (0.03) & 50.26 \\ 
   & deselection 0.1\% & 10 (0) & 10 (0) & 10 (0) & 10 (0) & 4.29 (1.33) & 3.82 (1.78) & 7.95 (2.94) & 3.49 (1.61) & 7.53 (2.72) & 1 (0) &  2647.89 (53.8) & 1.11 (0.03) & 61.82\\ 
   & deselection 1\% & 10 (0) & 5.18 (0.96) & 10 (0) & 4.76 (1.2) & 1.29 (1.03) & 1 (0) & 1.02 (0.14) & 1 (0) & 1 (0) & 1 (0) &  2933.74 (61.58) & 1.13 (0.03) & 60.78 \\ 
  & probing & 0.35 (0.97) & 4.4 (1.53) & 0.77 (1.42) & 4.22 (1.59) & 0 (0) & 0.16 (0.37) & 1.81 (1.31) & 0.31 (0.46) & 1.75 (1) & 0.52 (0.50) & 4205.81 (81.22) & 1.69 (0.04) & 1.49 \\ 
  & stabsel\_q20\_pfer5 & 0.12 (0.36) & 4.15 (1.06) & 0.96 (1.16) & 3.63 (1.03) & 0 (0) & 0 (0) & 0.17 (0.39) & 0 (0) & 0.21 (0.49) & 0 (0) &  4011.31 (88.88) & 1.55 (0.95) & 93.03 \\ \midrule
\multirow{5}{*}{\rotatebox[origin=c]{90}{Gauss}} &  classic & 10 (0) & 10 (0) & 10 (0) & 10 (0) & 10 (0) & 29.63 (6.41) & 19.51 (4.57) & 26.71 (6.42) & 20.36 (4.81) & 1.36 (0.59) & 2436.1 (56.94) & 1.09 (0.02) & 41.22 \\ 
   & deselect 0.001 & 10 (0) & 10 (0) & 10 (0) & 10 (0) & 10 (0) & 1.6 (0.72) & 3 (1.27) & 1.69 (0.8) & 3.12 (1.47) & 1.1 (0.33) & 2378.96 (49.84) & 1.09 (0.02) &  51.12 \\ 
   & deselect 0.01 & 10 (0) & 4.26 (0.58) & 10 (0) & 3.38 (0.63) & 8.63 (1.08) & 1 (0) & 1 (0) & 1 (0) & 1 (0) & 1 (0) & 2848.69 (59.78) & 1.12 (0.03) &  50.82 \\ 
   & probing & 1.2 (1.86) & 5.41 (1.52) & 2.16 (1.98) & 5.1 (1.67) & 0.34 (0.67) & 0.4 (0.49) & 1.39 (0.65) & 0.67 (0.47) & 1.4 (0.62) & 0.24 (0.43) & 4060.62 (122.73) & 1.64 (0.08) &  1.20  \\ 
 & stabsel\_q20\_pfer5 & 0.33 (0.59) & 4.69 (1.11) & 1.55 (1.19) & 4.11 (1.03) & 0.08 (0.27) & 0 (0) & 0.06 (0.24) & 0 (0) & 0.11 (0.31) & 0 (0) & 3970.84 (97.92) & 1.5 (1.04) & 61.94\\ \midrule

 \multirow{5}{*}{\rotatebox[origin=c]{90}{Gumbel}} & classic & 10 (0) & 10 (0) & 10 (0) & 10 (0) & 6.89 (1.45) & 46.02 (8.41) & 39.36 (7.07) & 42.12 (8.84) & 39.28 (6.41) & 1.09 (0.32) & 2485.89 (52.03) & 1.09 (0.03) &  88.28 \\ 
 & deselection 0.1\% & 10 (0) & 10 (0) & 10 (0) & 10 (0) & 6.19 (1.32) & 2.53 (1.34) & 5.85 (2.59) & 2.36 (1.16) & 5.33 (1.94) & 1.01 (0.1) & 2415.93 (52.52) & 1.09 (0.03) & 108.30 \\ 
   & deselection 1\% & 10 (0) & 4.97 (0.82) & 10 (0) & 4.29 (1.06) & 3.26 (0.81) & 1 (0) & 1 (0) & 1 (0) & 1 (0) & 1 (0) & 2809.8 (61.68) & 1.12 (0.03) &  107.45 \\ 
   & probing & 1.33 (2.38) & 4.97 (1.74) & 1.74 (2.32) & 4.26 (1.85) & 0 (0) & 0.31 (0.46) & 1.6 (1.24) & 0.51 (0.5) & 1.59 (0.96) & 1 (0) & 4130.46 (211.8) & 1.65 (0.10) & 6.79 \\
    & stabsel\_q20\_pfer5 & 0.19 (0.44) & 4.2 (0.98) & 1.44 (1.24) & 3.48 (1.04) & 0 (0) & 0 (0) & 0.14 (0.35) & 0 (0) & 0.14 (0.4) & 0 (0) & 3976.13 (95.14) & 1.42 (0.87) &  128.40  \\ 
   \bottomrule
\end{tabular}}
\caption{Low-dimensional setting with more informative predictors: Results of the classical boosted model, the deselection approach with $\tau =0.01$, stability selection and probing for the Clayton, Gaussian and Gumbel Copula in terms of mean (sd) of the true positives, false positives and the predictive performance in terms of negative log-likelihood and energy score (ES). The computational time corresponds to one simulation run. }
\end{table}

\begin{table}[t!]
\centering
\resizebox{\textwidth}{!}{%
\begin{tabular}{rlcccccccccccccccc}
  \toprule
  & & \multicolumn{5}{c}{True Positives}  & \multicolumn{5}{c}{False Positives} & & & &\\ 
 & method & $\mu_1$ & $\sigma_1$ & $\mu_2$ & $\sigma_2$ & $\rho$ & $\mu_1$ & $\sigma_1$ & $\mu_2$ & $\sigma_2$ & $\rho$ & -Log-Lik & ES & time\\ 
  \midrule
\multirow{4}{*}{\rotatebox[origin=c]{90}{Clayton}} & classic  & 10 (0) & 10 (0) & 10 (0) & 10 (0) & 0.34 (0.55) & 38 (11.75) & 60.67 (10.49) & 31.45 (11.41) & 58.49 (10.38) & 0.99 (0.10) & 2818.86 (58.88) & 1.12 (0.03) & 116.80 \\ 
  & deselection 0.1\% & 10 (0) & 10 (0) & 10 (0) & 10 (0) & 0.32 (0.55) & 4.79 (2.44) & 14.98 (3.88) & 3.87 (1.88) & 13.92 (3.85) & 1 (0) & 2770.18 (57.04) & 1.11 (0.03) & 124.11 \\ 
  & deselection 1\% & 10 (0) & 4.89 (0.9) & 10 (0) & 4.14 (1.02) & 0.05 (0.22) & 1 (0) & 1.02 (0.14) & 1 (0) & 1 (0) & 1 (0) & 2965.47 (60.16) & 1.13 (0.03) & 124.27 \\ 
   & probing & 0.07 (0.41) & 3.44 (1.27) & 0.11 (0.6) & 2.92 (1.34) & 0 (0) & 0.03 (0.17) & 1.61 (0.95) & 0.04 (0.20) & 1.68 (1.07) & 0.19 (0.39) & 4253.21 (59.46) & 1.71 (0.03) & 5.20 \\ \midrule
\multirow{4}{*}{\rotatebox[origin=c]{90}{Gauss}}  
 & classic & 10 (0) & 10 (0) & 10 (0) & 10 (0) & 10 (0) & 38.40 (9.64) & 36.43 (7.72) & 34.65 (8.8) & 34.85 (7.12) & 1.36 (0.63) & 2546.15 (56.01) & 1.10 (0.03) & 118.77 \\ 
   & deselection 0.01\% & 10 (0) & 10 (0) & 10 (0) & 10 (0) & 10 (0) & 2.29 (1.12) & 5.30 (2.13) & 2.48 (1.19) & 5.44 (2.52) & 1.08 (0.27) & 2456.10 (53.64) & 1.10 (0.03) & 127.97\\ 
   & deselection 1\% & 10 (0) & 4.24 (0.43) & 10 (0) & 3.31 (0.54) & 6.69 (1.25) & 1 (0) & 1 (0) & 1 (0) & 1 (0) & 1 (0) & 2883.14 (55.43) & 1.13 (0.03) & 127.52 \\ 
   & probing & 0.21 (1.01) & 4.29 (1.34) & 0.36 (1.31) & 3.57 (1.64) & 0.05 (0.41) & 0.07 (0.26) & 1.53 (0.92) & 0.13 (0.34) & 1.51 (1.08) & 0.02 (0.14) & 4147.48 (89.63) & 1.70 (0.06) & 4.30 \\  \midrule

 \multirow{4}{*}{\rotatebox[origin=c]{90}{Gumbel}} & classic & 10 (0) & 10 (0) & 10 (0) & 10 (0) & 2.51 (1.71) & 61.53 (19.35) & 60.64 (10.64) & 52.98 (17.08) & 58.69 (9.43) & 1 (0) & 2633.83 (59.71) & 1.11 (0.03) & 149.68 \\ 
 & deselection 0.1\% & 10 (0) & 10 (0) & 10 (0) & 10 (0) & 2.39 (1.68) & 4.09 (1.63) & 10.87 (3.76) & 3.85 (1.73) & 10.05 (3.27) & 1 (0) & 2543.5 (72.51) & 1.10 (0.03) &  167.55 \\ 
   & deselection 1\% & 10 (0) & 4.81 (0.8) & 10 (0) & 3.98 (0.9) & 0.75 (0.96) & 1 (0) & 1 (0) & 1 (0) & 1 (0) & 1 (0) & 2852.34 (62.38) & 1.12 (0.03) & 166.85 \\ 
   & probing & 0.07 (0.36) & 3.51 (1.29) & 0.27 (0.78) & 2.89 (1.34) & 0 (0) & 0.05 (0.22) & 1.63 (1.33) & 0.15 (0.36) & 1.31 (0.60) & 0.98 (0.14) & 4254.05 (73.17) & 1.71 (0.03) & 5.31 \\ 
   \bottomrule
\end{tabular}}
\caption{High-dimensional setting with more informative predictors: Results of the classical boosted model, the deselection approach with $\tau =0.01$, stability selection and probing for the Clayton, Gaussian and Gumbel Copula in terms of mean (sd) the true positives, false positives and the predictive performance in terms of negative log-likelihood and energy score (ES). The computational time corresponds to one simulation run. }
\end{table}

\newpage
\section{Analysis of fetal ultrasound data}
\nopagebreak
\begin{table}[h]
\tiny
    \centering
    \begin{adjustbox}{angle=90}
    \begin{tabular}{l|ccccc|ccccc|ccccc|ccccc|ccccc}
    \toprule
    & \multicolumn{5}{c}{classic} & \multicolumn{5}{c}{deselection $0.1\%$} &  \multicolumn{5}{c}{deselection $1\%$} &  \multicolumn{5}{c}{stability selection}  &  \multicolumn{5}{c}{probing}\\
    Variable & $\mu_1$ & $\sigma_1$ & $\mu_2$ & $\sigma_2$ & $\rho$ & $\mu_1$ & $\sigma_1$ & $\mu_2$ & $\sigma_2$ & $\rho$ &  $\mu_1$ & $\sigma_1$ & $\mu_2$ & $\sigma_2$& $\rho$ & $\mu_1$ & $\sigma_1$ & $\mu_2$ & $\sigma_2$& $\rho$ & $\mu_1$ & $\sigma_1$ & $\mu_2$ & $\sigma_2$& $\rho$ \\
    \midrule
    Abdominal sagittal diameter (APAD) & + & &  +  & & & & & & &&&&&&&&&&&&&&&&\\ 
    Abdominal transverse diameter (ATD) & + &  & + & & & & & &&&&&&&&&&&&&&&&&\\ 
    Abdominal circumference (AC) &  + & & +  & & & & & &&&&&&&&&&&&&&&&&\\ 
    Biparietal diameter (BPD) &  + & + & & & & &&& &&&&&&&&&&&&&&&\\
    Occipitofrontal diameter (OFD) & + & & & & & & &&&&&&&&&&&&&&&&&\\
    Head circumference (HC) & + & + & + & + &&& & & &  & &&&&&&&&&&&&&& \\
    Femur length (FL) & + & & +  & & & & & &&&&& &&&& &&&&&&&&\\ 
    
    APAD*ATD & + &  & +  & & +  & & & + &&&&&&&&&&&&&&&&&+\\
    APAD*AC & + & & + & & +  & & & &&&&&&&&&&&&&&&&& \\
    APAD*BPD & + &  + & +  & +  & & & & & +& &&&&&&&&&& &&&& +&\\
    APAD*OFD & + &  & + & +  & & & & & &&&&&&&&&&&&&&& +&\\
    APAD*HC & + & + & & + & & & + &  & +& &&+&&&&&&&&&& +& & + &\\
    APAD*FL & + & + & + & & + & & + & && +&&+&&&&& + &&&&& +&&&+\\
    
    ATD*AC & + & & + & & & &  & & &&&&&&&&&& &&&&&&\\
    ATD*BPD & + & & + & +  & & & & & & &&&&&&&&&&&&&&&\\
    ATD*OFD & + & + & +  & & & & + & && && &&&&&&&&&& +&&& \\
    ATD*HC &  &  & & & &  && &&&&&&&&&&&&&&\\
    ATD*FL & + & + & + & & +  & & & && +& &&& & + & &&&&& + && +& &+\\
    
    AC*BPD & + & + & +  & + & & + & +  & + & +& & & & + & + & & +&& +& + && +& +& +& +&\\
    AC*OFD & + &  & +  & & &  & & & && & && &&&& &&&&&&&\\
    AC*HC & + & + & +  &+ & & & + & + & +& & &+ && +&&&& + & && +& +& +& + & \\
    AC*FL & + & + & +  & + & +  & + & + & + & +& + && + & + & + & & + & +& +& + && +& + & + & + &+\\
    
    BPD*OFD & + &  & +  & & & & & & & &&&&&&&&&&&&&&& \\
    BPD*HC & + & & & & &  & & & & && &&&&&&&&& &&&&\\
    BPD*FL & + & & + & & & +& && & & &&&&& + &&& && +&&&& \\
    OFD*HC &  &  & +  & & & & & & &&&&&&&&&& &&&&&&\\
    OFD*FL & + & & + & + & +  & & & & & + &&&&&&&&&&&&&&&+ \\
    
    HC*FL & + & &  + & & & + & & + & & &&&&&& + &&&&& +&&+& &\\
    Weight (mother) & + & + & +  & + & + & & + &+ & &&&&&&&&&&&&& +& + & + &\\
    Height (mother) & + & + & + & & +  & & + & & && &&&&&&&&&& + & +& &&\\
    BMI (mother) & + & + & + & + & && & & + & &&& &&&&&&&&&&&+&\\
    Gravida & + & + & +  & + & & & & & & &&&&&&&&&&&&&&&\\
    Para & + & & + & & & & & & &&&&&&&&&&&&&& +&&\\
    Gestational age & + & + & +  & + & + & + & + & + & + & && +&&+& & + & +&& + & & + & +& +& + &\\
    Gestational diabetes &  &  & + & & & &  & & & & && &&&&&&&&&&&\\
    Sex & + & & + & & & + & & + &&&&&&&&&&&&& +&&&&\\
         \bottomrule
    \end{tabular}      \end{adjustbox}
    \caption{Selected covariates of the fetal ultrasound data for the classical boosted model, the deselection approach with 0.1\% and 1\%, stability selection and probing.}

\end{table}
\end{document}